\documentclass[prd,aps,floatfix,preprint,nofootinbib]{revtex4}
\usepackage{graphicx}
\usepackage{amsmath}
\usepackage{amssymb}
\usepackage{longtable}
\usepackage{axodraw}

\def\drvstar#1{\partial\kern-0.5pt\smash{\raise 4.5pt\hbox{$\ast$}}
               \kern-5.0pt_{#1}} 
\def\lvec#1{\setbox0=\hbox{$#1$}
    \setbox1=\hbox{$\scriptstyle\leftarrow$}
    #1\kern-\wd0\smash{
    \raise\ht0\hbox{$\raise1pt\hbox{$\scriptstyle\leftarrow$}$}}
    \kern-\wd1\kern\wd0} 
 
\def\ldrvstar#1{\lvec{\,\partial}\kern-0.5pt\smash{\raise 4.5pt\hbox{$\ast$}}
               \kern-5.0pt_{#1}}
\newcommand{\simg}{\rlap{\raise -4pt \hbox{$\sim$}}
                   \raise 3pt \hbox{$>$}}
\newcommand{\siml}{\rlap{\raise -4pt \hbox{$\sim$}}
                   \raise 3pt \hbox{$<$}}
  
\newcommand{\no}{\nonumber}

\begin{document}   
 \vspace*{-20mm}
 \begin{flushright}
  \normalsize
  KEK-CP-205\\
  UTHEP-555\\
  YITP-07-85
 \end{flushright}
\title{
 $B_K$ with two flavors of dynamical overlap fermions
}
\author{
 S.~Aoki$^{a,b}$, H.~Fukaya$^{c}$, S.~Hashimoto$^{d,e}$,
 J.~Noaki$^{d}$, T.~Kaneko$^{d,e}$, H.~Matsufuru$^d$,  T.~Onogi$^f$,
 N.~Yamada$^{d,e}$\\[1ex]
 (JLQCD Collaboration)
}
\affiliation{
$^a$Graduate School of Pure and Applied Sciences, University of
 Tsukuba, Tsukuba 305-8571, Japan\\
$^b$Riken BNL Research Center, Brookhaven National Laboratory, Upton,
    New York 11973, USA\\
$^c$The Niels Bohr Institute, The Niels Bohr International Academy,
 Blegdamsvej 17 DK-2100 Copenhagen {\O}, Denmark\\
$^d$High Energy Accelerator Research Organization (KEK), Tsukuba
    305-0801,Japan\\
$^e$School of High Energy Accelerator Science, The Graduate University
    for Advanced Studies (Sokendai), Tsukuba 305-0801, Japan\\
$^f$Yukawa Institute for Theoretical Physics, Kyoto University,
           Kyoto 606-8502, Japan
}
\date{\today}
\begin{abstract}
 We present a two-flavor QCD calculation of $B_K$ on a $16^3 \times 32$
 lattice at $a\sim 0.12$ fm (or equivalently  $a^{-1}$=1.67 GeV).
 Both valence and sea quarks are described by the overlap fermion
 formulation.
 The matching factor is calculated nonperturbatively with the
 so-called RI/MOM scheme.
 We find that the lattice data are well described by the next-to-leading
 order (NLO) partially quenched chiral perturbation theory (PQChPT) up
 to around a half of the strange quark mass ($m_s^{\rm phys}/2$).
 The data at quark masses heavier than $m_s^{\rm phys}/2$ are fitted
 including a part of next-to-next-to-leading order terms.
 We obtain
 $B_K^{\overline{\rm MS}}(2\ {\rm GeV})= 0.537(4)(40)$, where the
 first error is statistical and the second is an estimate of systematic
 uncertainties from finite volume, fixing topology, the matching factor,
 and the scale setting.
\end{abstract}
\maketitle
\setcounter{footnote}{0}

\section{Introduction}
\label{sec:introduction}

The indirect CP violation in neutral kaon decays, characterized by the
$|\epsilon_K|$ parameter, plays an important role to constrain the
Cabibbo-Kobayashi-Maskawa (CKM) matrix elements.
Experimentally, $|\epsilon_K|$ has been measured to an excellent precision,
$|\epsilon_K| = (2.233 \pm 0.015)\times 10^{-3}$~\cite{Yao:2006px},
through the two pion decays of long-lived neutral kaons.
Within the standard model, $|\epsilon_K|$ is expressed
as~\cite{Buchalla:1995vs}
\begin{eqnarray}
 |\epsilon_K|=(\mbox{known factor})\times B_K(\mu) \times
           f(\bar\rho,\bar\eta),
\label{eq:epsilon}
\end{eqnarray}
where $f(\bar\rho,\bar\eta)$ is a known function of the Wolfenstein
parameters, $\bar\rho$ and $\bar\eta$, and $B_K(\mu)$ is a less known
hadronic matrix element defined by
\begin{eqnarray}
   B_K(\mu)
 = \frac{\langle \overline{K^0} |\,
    \bar d \gamma_\mu(1-\gamma_5)s\
    \bar d \gamma_\mu(1-\gamma_5)s\,
   | K^0 \rangle}
   {\frac{8}{3}f_K^2 m_K^2}.
   \label{eq:bk definition}
\end{eqnarray}
The $\Delta S = 2$ four-quark operator
$\bar{d}\gamma_\mu(1-\gamma_5)s \bar{d}\gamma_\mu(1-\gamma_5)s$ is
renormalized at a scale $\mu$.
The parameters appearing in the denominator, $f_K$ and $m_K$, denote the
leptonic decay constant and mass of kaon, respectively.
Since the precision of the constraint on the fundamental parameters such as
$(\bar{\rho},\bar{\eta})$ is limited by the theoretical uncertainty in the
kaon $B$ parameter $B_K$, its precise calculation has a direct relevance to
the study of the flavor structure of the standard model and beyond.
The purpose of this study is to provide such a precise calculation of $B_K$
using lattice QCD. 

The numerator in~(\ref{eq:bk definition}) involving the
$(V-A)\times (V-A)$ four-quark operator behaves as $m_K^2$ for the
external states at rest, hence it vanishes in the chiral limit of kaon.
This behavior is altered if the four-quark operator mixes with other
operators with different chiral structures under renormalization.
Since the denominator in~(\ref{eq:bk definition}) contains $m_K^2$ and
so vanishes in the chiral limit too, the appearance of operators with
different chirality in the numerator causes unphysical divergence in the
ratio toward the chiral limit.
In the lattice calculation with the Wilson-type fermion formulations,
this problem occurs because the formulations explicitly violate the
chiral symmetry.
Calculation of $B_K$ using the Wilson-type fermions, therefore, is not
very precise due to the uncontrolled operator mixing (for recent efforts
in unquenched calculations, see~\cite{Flynn:2004au,Mescia:2005ew}).
For this reason, the lattice calculation of $B_K$ has historically been
done using the staggered fermion formulation, with which the chiral
symmetry is realized at a cost of introducing unwanted flavors.
A quenched calculation using the staggered fermion~\cite{Aoki:1997nr}
has long been a ``benchmark'' calculation until recently,
which is $B_K(2\mathrm{~GeV}) = 0.628(42)$ without an estimate of
quenching uncertainty. 
More recently, the domain-wall fermion, which respects an approximate
chiral symmetry on the lattice without introducing extra flavors, is
applied to the calculation of $B_K$.
Quenched calculations showed that the lattice artifact is significantly
smaller than that of the staggered fermion and hence the continuum
extrapolation is more
straightforward~\cite{AliKhan:2001wr,Blum:2001xb,Aoki:2005ga}.
Therefore, a great effort has been made to realize unquenched
simulations using the domain-wall fermion; a 2+1-flavor calculation
result has recently been presented by the RBC-UKQCD
collaboration~\cite{Antonio:2007pb}, which is
$B_K(2\mathrm{~GeV}) = 0.524(10)(28)$ with a combined statistical (the
first error) and systematic (the second) error being 6\%.
(An earlier result of two-flavor QCD is also
available~\cite{Aoki:2004ht}.)

This work improves these lattice calculations of $B_K$ in several
directions.
First of all, we use the overlap fermion formulation, which exactly
respects a lattice variant of chiral symmetry.
The problem of the operator mixing is absent in this formulation.
With the domain-wall fermion, this is a nontrivial problem, because
there is a tiny but nonzero operator mixing which could be enhanced by
the unsuppressed chiral behavior of wrong chirality operators as
discussed above.
A detailed study gives an estimate of order 0.1\% effect for
$B_K$~\cite{Aoki:2007xm}, which is negligible in the precision we are
aiming at, but a calculation without such a delicate problem from the
beginning is desirable.
The exact chiral symmetry also helps to further reduce the
discretization effect, since the $O(a)$ effect is completely absent.

Second, the use of chiral perturbation theory (ChPT), and even partially
quenched (PQ) ChPT is justified in the analysis of lattice
data, especially in the chiral extrapolation.
Because of the pion (and kaon) loop effect, $B_K$ develops a nonanalytic
quark mass dependence, the so-called chiral logarithm, which may give a
non-negligible contribution in the chiral extrapolation of lattice data
to the physical up and down quark masses. 
Whether the observed quark mass dependence of lattice data is well
described by ChPT is a complicated problem, since the mass region where
the ChPT formula is applicable is not known from the outset and thus has
to be tested by the lattice calculation for each quantity of interest.
This test is difficult without the exact chiral symmetry, because the
ChPT formula itself must be modified by including the effect of the
violated chiral symmetry.
Another requirement for an unambiguous test is a sufficient number of
data points.
We explore a broader sea and valence quark mass region covered with
significantly more data points than former studies.
In this work, we perform the test using the next-to-leading order (NLO)
PQChPT formula by varying the sea quark mass region in the fit.
The statistical signal of individual data points is improved by accumulating more
statistics and by using a new technique, {\it i.e.}, low-mode averaging.
As a consequence, we are able to identify the applicability region of the NLO
PQChPT, which makes the chiral extrapolation more reliable.

In spite of these obvious advantages, the overlap fermion formulation
has not been used extensively especially for dynamical fermion
simulations.
The main problem is in its large computational cost to approximate the
matrix sign function appearing in the definition of the overlap
operator.
Superficially, the cost is similar to that of the domain-wall fermion which
requires $N_s$ (length in the fifth dimension) times more computation,
but in the hybrid Monte Carlo (HMC) simulation, the overlap fermion is much
harder because of the discontinuity of the sign function, which requires
a special trick, such as the
``reflection-refraction''~\cite{Fodor:2003bh}.
This makes the dynamical overlap fermion simulation substantially more
costly.
In this work, we avoid this problem by introducing a topology fixing
term to a gauge field action~\cite{Fukaya:2006vs}, with which we never
encounter the discontinuity of the sign function.
The physical effect of fixing the topological charge can be understood
and be estimated, at a solid theoretical ground, as a finite volume
effect~\cite{Aoki:2007ka}.

The overlap fermion simulation has already been applied to a calculation
of pion and kaon masses and decay constants~\cite{Noaki:2007es}, pion
form factor~\cite{Kaneko:2007nf}, $\pi^\pm-\pi^0$ mass
splitting~\cite{Shintani:2007ub}, topological
susceptibility~\cite{Aoki:2007pw}, and more applications are planned.
The simulation is also extended to the $\epsilon$-regime, where the sea
quark mass is even smaller than the physical value, and is used to
extract the chiral condensate and pion decay constant in the chiral
limit~\cite{Fukaya:2007fb,Fukaya:2007yv,Fukaya:2007cw,Fukaya:2007pn}.
An overview of our project is presented in~\cite{Matsufuru:2007uc}.

A potential problem of our work is that the simulation is performed with
two flavors of sea quarks that correspond to up and down quarks.
The effect of the strange sea quark is neglected.
Although we do not expect its significant effect on $B_K$, the actual
correction is hardly estimated within the two-flavor theory.
We therefore have a plan to extend this work to a 2+1-flavor
calculation~\cite{Hashimoto:2007vv}, for which this work serves as a
prototype calculation with an almost realistic setup. 

The rest of this paper is organized as follows.
The lattice actions and simulation parameters are described in
Sec.~\ref{sec:parameters}.
The methods to calculate the bare $B_K$ and the nonperturbative
matching factor are introduced in Sec.~\ref{sec:Method and results}.
In Sec.~\ref{sec:test-PQChPT}, we compare the quark mass dependence of
$B_K$ with the prediction of NLO PQChPT to see the consistency.
In Sec.~\ref{sec:extraction of BK}, how to extract the physical $B_K$ is
presented.
Systematic errors in our result are discussed in
Sec.~\ref{sec:systematic errors}.
A summary of this work is given in Sec.~\ref{summary}.

\section{Simulation parameters}
\label{sec:parameters}

We perform the calculation on a $16^3\times 32$ lattice using the
Iwasaki gauge action.
The periodic boundary condition is set in all four directions.
Both dynamical and valence quarks are described by the overlap-Dirac
operator~\cite{Neuberger:1997fp,Neuberger:1998wv},
\begin{eqnarray}
  D_{\rm ov}(m_q)
= \left(m_0+\frac{m_q}{2}\right)
 +\left(m_0-\frac{m_q}{2}\right)\gamma_5\,{\rm sgn}(H_W(-m_0)),
\end{eqnarray}
where $m_q$ is a quark mass and $H_W(-m_0)$ denotes the standard
Hermitian Wilson-Dirac operator with a negative mass.
Throughout this work we take $m_0$=1.6.
A generation of configurations with dynamical overlap quarks requires a huge
computational cost.
To accelerate HMC, we introduce extra (unphysical) Wilson fermion and
ghost fields, which suppress the appearance of the zero mode of
$H_W(-m_0)$~\cite{Fukaya:2006vs}.
At a price, the global topological charge $Q$ is frozen during the HMC
evolution.
At $\beta$=2.30 the inverse lattice spacing is $1/a=1.67(2)(2)$ GeV,
which is determined through $r_0=0.49$ fm~\cite{Sommer:1993ce} in the
$Q=0$ sector.
The physical spatial volume of our lattice is about (1.9 fm)$^3$.
We carry out the simulations at six sea quark masses:
$m_{\rm sea}$=0.015, 0.025, 0.035, 0.050, 0.070, 0.100 in the lattice
unit.
These approximately cover the mass range
$[m_s^{\rm phys}/6,~m_s^{\rm phys}]$, where $m_s^{\rm phys}$ denotes the
physical strange quark mass.
The lightest pion is about 290 MeV, which gives $m_\pi L\sim$2.7.
The main calculation is made in the $Q=0$ sector.
In order to study the topological charge dependence, we have also
generated configurations in the $Q=-2$ and $-4$ sectors at
$m_{\rm sea}$=0.050.
We have accumulated 10000 trajectories for each sea quark mass at $Q=0$
and 5000 at $Q=-2$ and $-4$.
More details about the algorithm and parameters to generate the
configurations are described in~\cite{Kaneko:2006pa}.

The calculation of $B_K$ is done at every 20 trajectories for each sea quark
mass, and a single jackknife bin consists of 5 measurements.
For each sea quark mass, we take six valence quark masses and calculate
$B_K$ with all possible combinations of two valence quarks.
The gauge is fixed to the Coulomb gauge except for the calculation of
the nonperturbative matching factor, which is done in the Landau
gauge.
Low-mode averaging~\cite{DeGrand:2004qw} is implemented for all
correlation functions, which substantially improves the statistical
signal.

\section{Method and results}
\label{sec:Method and results}

\subsection{Two-point functions and pseudoscalar meson masses}
\label{subsec:2pt}

\begin{figure}
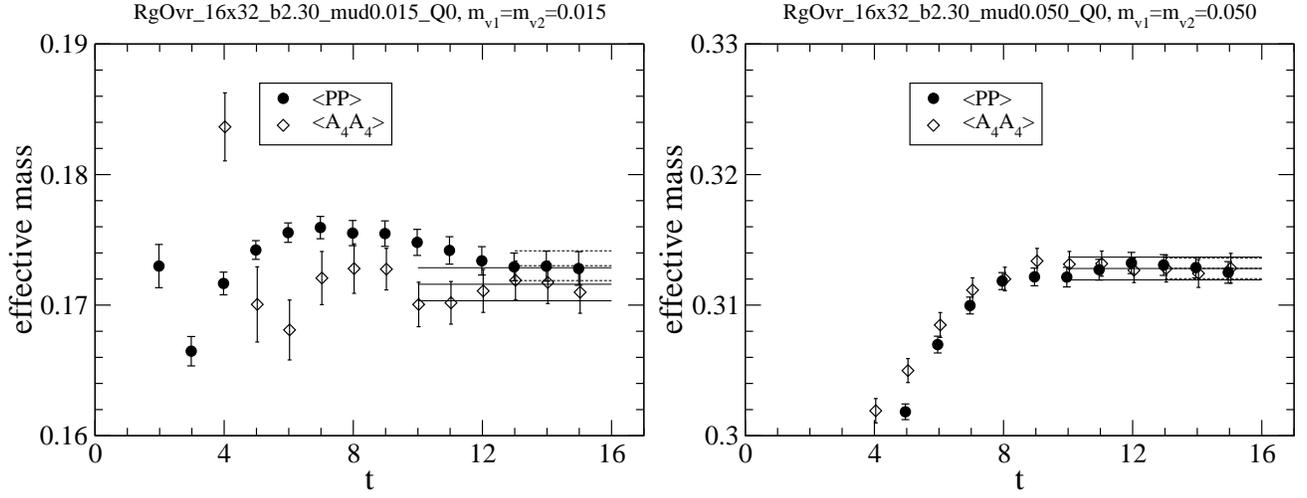

 \centering
 \begin{tabular}{cc}
  \includegraphics*[width=0.5 \textwidth,clip=true]
  {figs/efp_m0.015_001001.eps} &
  \includegraphics*[width=0.5 \textwidth,clip=true]
  {figs/efp_m0.050_004004.eps}
 \end{tabular}
 \caption{Effective mass plots for the pseudoscalar meson with
 $m_{\rm sea}=m_{v1}=m_{v2}$.
 The plots for $m_{\rm sea}$=0.015 (left) and 0.050 (right) are shown
 as representatives.}
 \label{fig:efplt}
\end{figure}

We calculate kaon two-point functions
\begin{eqnarray}
   C^{(2),\rm p-w}_{A_4A_4}(t)
 = \sum_{\vec x}\langle\,A_4(t,\vec x)\,
   A_4^{\rm wall}(0)\,\rangle.
 \label{eq:two-pt-def}
\end{eqnarray}
with a wall source at $t_{\rm src}$ and a point sink at $t+t_{\rm src}$.
$A_4(t,\vec x)= \bar q_1(t,\vec x)\,\gamma_4\gamma_5\,q_2'(t,\vec x)$ is
the axial-vector current.
The wall source is defined by
$A^{\rm wall}_4(t)=\big(\sum_{\vec x} \bar q_2(t,\vec x)\big)\,
\gamma_4\gamma_5\,\big(\sum_{\vec y}q_1(t,\vec y)\big)$.
$q_1$ and $q_2$ represent two different flavors of quarks described by
the overlap formalism, and in the definition of the axial-vector current
$q_2$ is modified to
$q_2'(x)=\left[1-D_{\rm ov}(0)/(2\,m_0)\right]q_2(x)$ such that the
axial-vector current exactly transforms into the vector current
$V_4(t,\vec x)= \bar q_1(t,\vec x)\,\gamma_4\,q_2'(t,\vec x)$ under the
axial transformation.
We take an average of physically equivalent two-point functions over the
four source points $t_{\rm src}=0, 8, 16, 24$.
Because of the periodic boundary condition in the time direction, its
asymptotic behavior in large $t$ is given by
\begin{eqnarray}
 \mbox{eq.~(\ref{eq:two-pt-def})}
 \rightarrow
     \frac{V_3\,Z^{\rm wall}_{A_4}}{2\,m_P}
     f_P\,m_P\,
     \left(e^{-m_P\,t}+e^{m_P\,(t-N_t)}
     \right),
     \label{eq:2-pt}
\end{eqnarray}
where $N_t=32$ and $V_3=16^3$.
Data at two time slices equally separated from $t=16$ are averaged.
The correlated fit is carried out to extract $m_P$ and
$Z^{\rm wall}_{A_4}f_P/2$, where
\begin{eqnarray}
     Z^{\rm wall}_{A_4}
 &=& \langle P|\,\sum_{\vec x}
     \bar q_2(0,\vec x)\,\gamma_4\gamma_5\,q_1(0,\vec 0)\,
     |\,0\rangle,\\
     f_P\,m_P
 &=& \langle0|A_4(0)|P\rangle.
\end{eqnarray}
The fit range is chosen to be $t=[\,9,\,16]$ for all the sea and valence
quark masses.
The numerical results for $m_P$ and $Z_{A_4}^{\rm wall}f_P/2$ are listed
in Tables~\ref{tab:mass-results-0.015}-\ref{tab:mass-results-0.050-Q-4}
for each ensemble.
We also calculate pseudoscalar two-point functions $\langle PP\rangle$
and make a correlated fit to obtain the mass.
Effective mass plots for the pseudoscalar and axial-vector two-point
functions are compared in Fig.~\ref{fig:efplt}.
Horizontal lines show the mass and one-sigma band obtained from the fit,
also indicating the fit ranges chosen.
We confirm that the masses extracted from $\langle PP\rangle$ and
$\langle A_4A_4\rangle$ are consistent within 1 standard deviation for
all the cases.
We use the mass from $\langle A_4A_4\rangle$ in the following
analysis.

\subsection{Three-point functions and $B_P$}
\label{subsec:3pt}

\begin{figure}
 \centering
 \includegraphics*[width=0.6 \textwidth,clip=true]
  {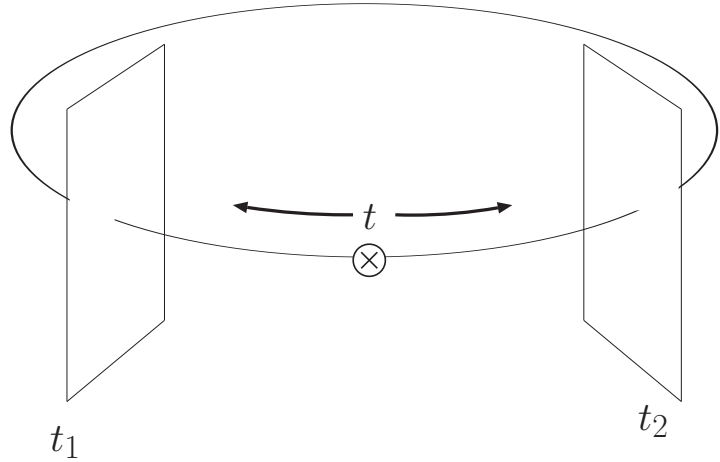}
 \caption{The setup of two wall sources at $t_1$ and $t_2$ and the local
 operator at $t$ to calculate the three-point function.}
 \label{fig:method}
\end{figure}

In order to obtain $B_K$ we also need to calculate three-point
functions defined by
\begin{eqnarray}
 \hspace*{-3ex}
   C^{(3)}_{L_\mu L_\mu}(t_2,t,t_1)
 = \sum_{\vec x} \langle\,
   A_4^{\rm wall}(t_2)\
   O^{\rm lat}_{L_\mu L_\mu}(t,\vec x)\
   A^{\rm wall}_4(t_1)\,\rangle,
 \label{eq:3pt-def}
\end{eqnarray}
where
\begin{eqnarray}
 O^{\rm lat}_{L_\mu L_\mu}
=\bar q_1 \gamma_\mu(1-\gamma_5)q_2'\
 \bar q_1 \gamma_\mu(1-\gamma_5)q_2'
\label{eq:four-quark-op}
\end{eqnarray}
is the $\Delta S$=2 four-quark operator defined on the lattice.
In the calculation of three-point functions, the kaon interpolating
operator, that is the wall source, is put at fixed time slices $t_1$ and
$t_2$ while the position of four-quark operator $t$ is varied as shown
in Fig.~\ref{fig:method}.
We take $|t_2-t_1|$=16 or 24, which we call set A and B respectively.
For set A we take $(t_2,t_1)$=(16, 0) and (24, 8), while for set B we take
$(t_2,t_1)$=(24, 0), (32, 8), (8, 16), and (16, 24).
Within each set, all the three-point functions are equivalent after
proper translation in the time direction, so they are averaged after
the translation.
For the set A (with $|t_2-t_1|$=16), two equivalent regions, $0<t<16$
and $16<t<32$ are further averaged.
The averaged three-point functions are finally shifted such that the
wall source for the set A and B is $(t_2, t_1)=(16,\ 0)$ and $(32,\ 8)$,
respectively.

\begin{figure}
 \centering
 \includegraphics*[width=0.6 \textwidth,clip=true]
  {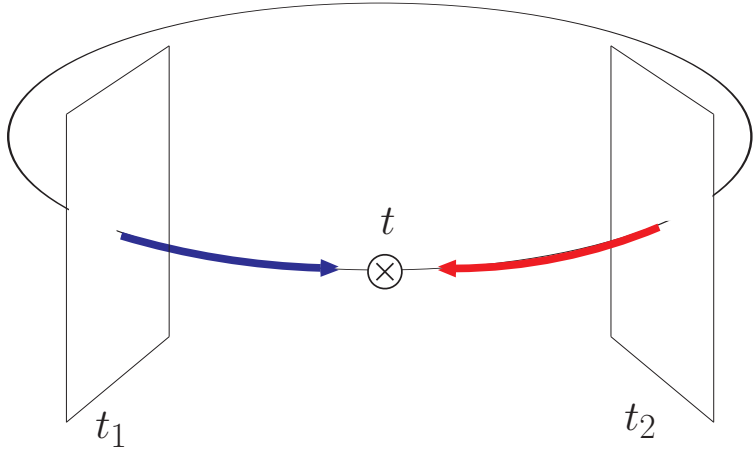}
 \caption{The contribution of the three-point function relevant to
 the calculation of $B_K$.}
 \label{fig:relevant}
\end{figure}

\begin{figure}
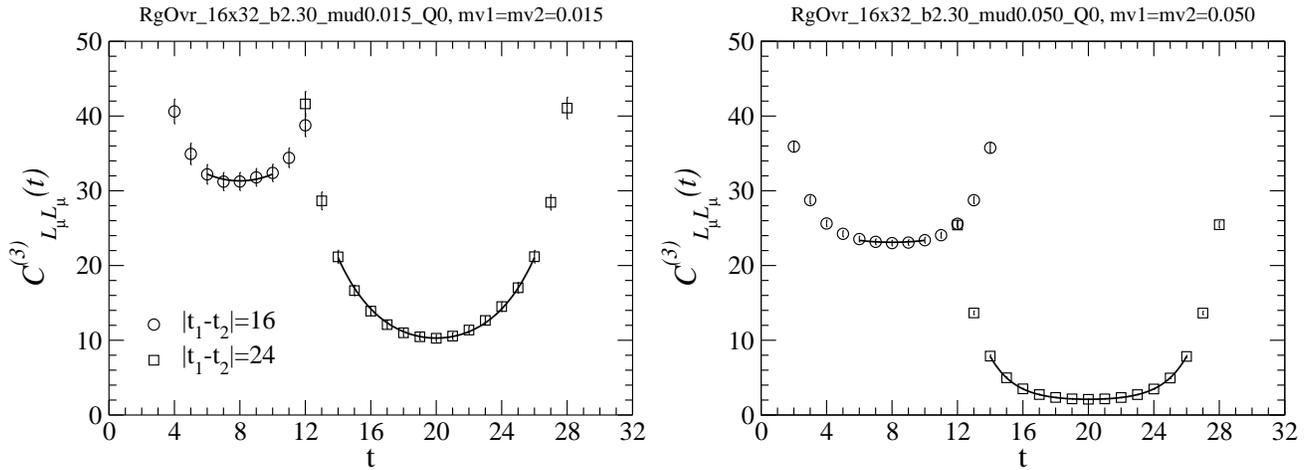

 \centering
 \includegraphics*[width=0.5 \textwidth,clip=true]
 {figs/3pt_LL_m0.015_001001.eps}~
 \includegraphics*[width=0.5 \textwidth,clip=true]
 {figs/3pt_LL_m0.050_004004.eps}
 \caption{$t$-dependence of the three-point functions.
 The left is for $m_{\rm sea}=m_{v1}=m_{v2}=0.015$, and the right for
 0.050.
 The fit ranges are $[t_{\rm min},t_{\rm max}]=[6,10]$ and $[14,26]$
 for the set A and B, respectively.}
 \label{fig:3pt}
\end{figure}

\begin{figure}
 \centering
 \begin{tabular}{cc}
 \includegraphics*[width=0.4 \textwidth,clip=true]
  {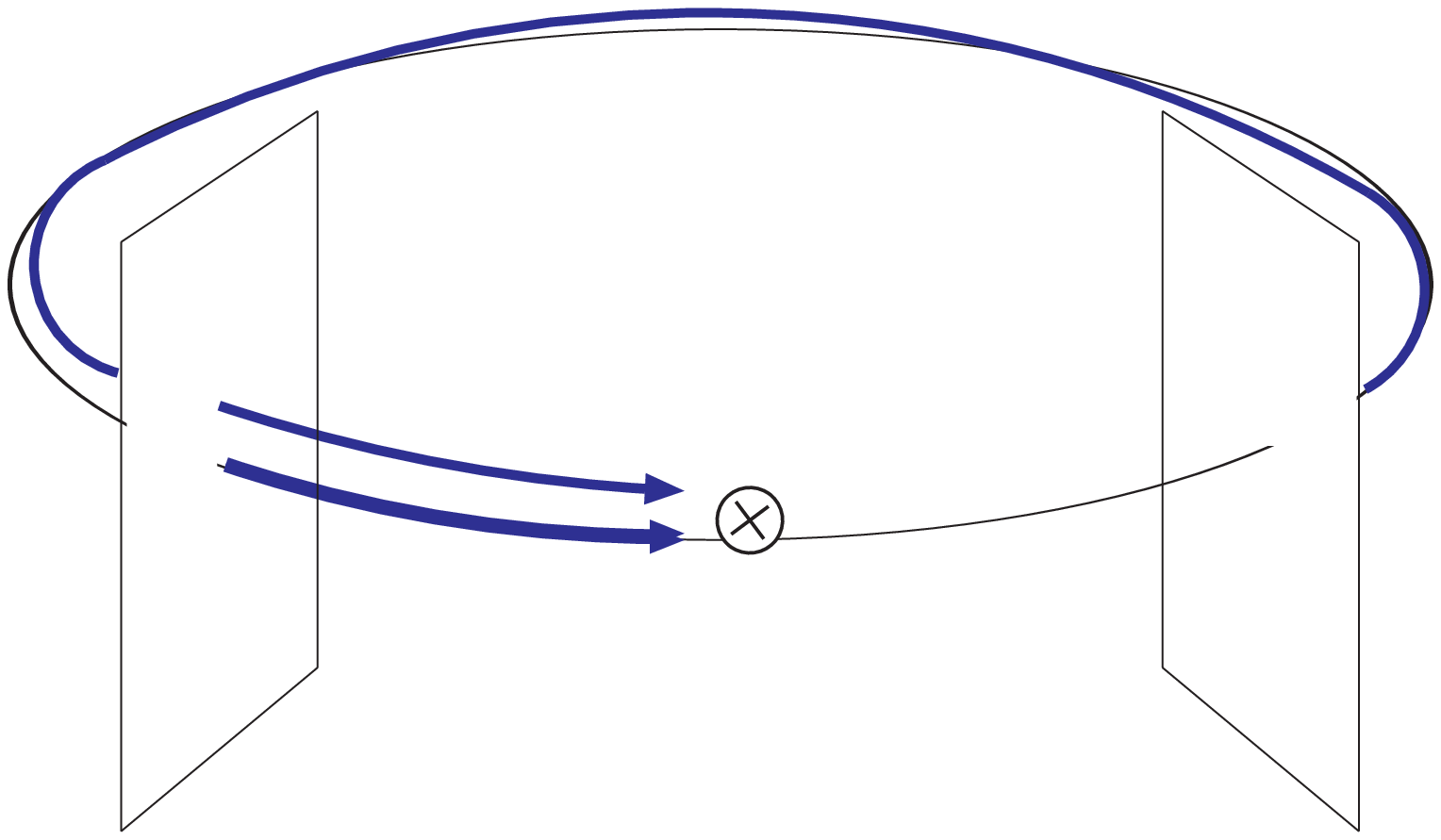}&
 \includegraphics*[width=0.4 \textwidth,clip=true]
  {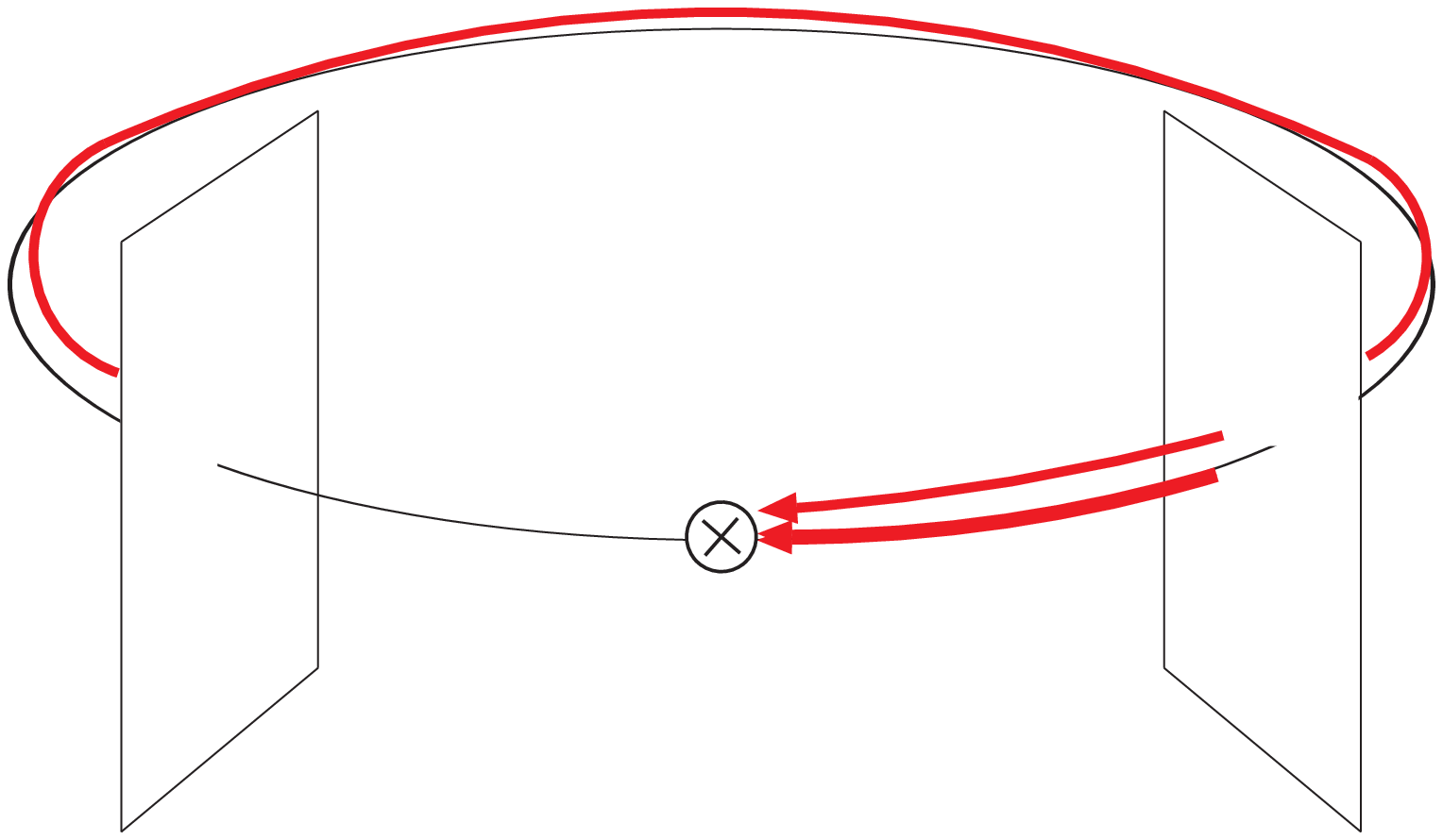}\\
 \includegraphics*[width=0.4 \textwidth,clip=true]
  {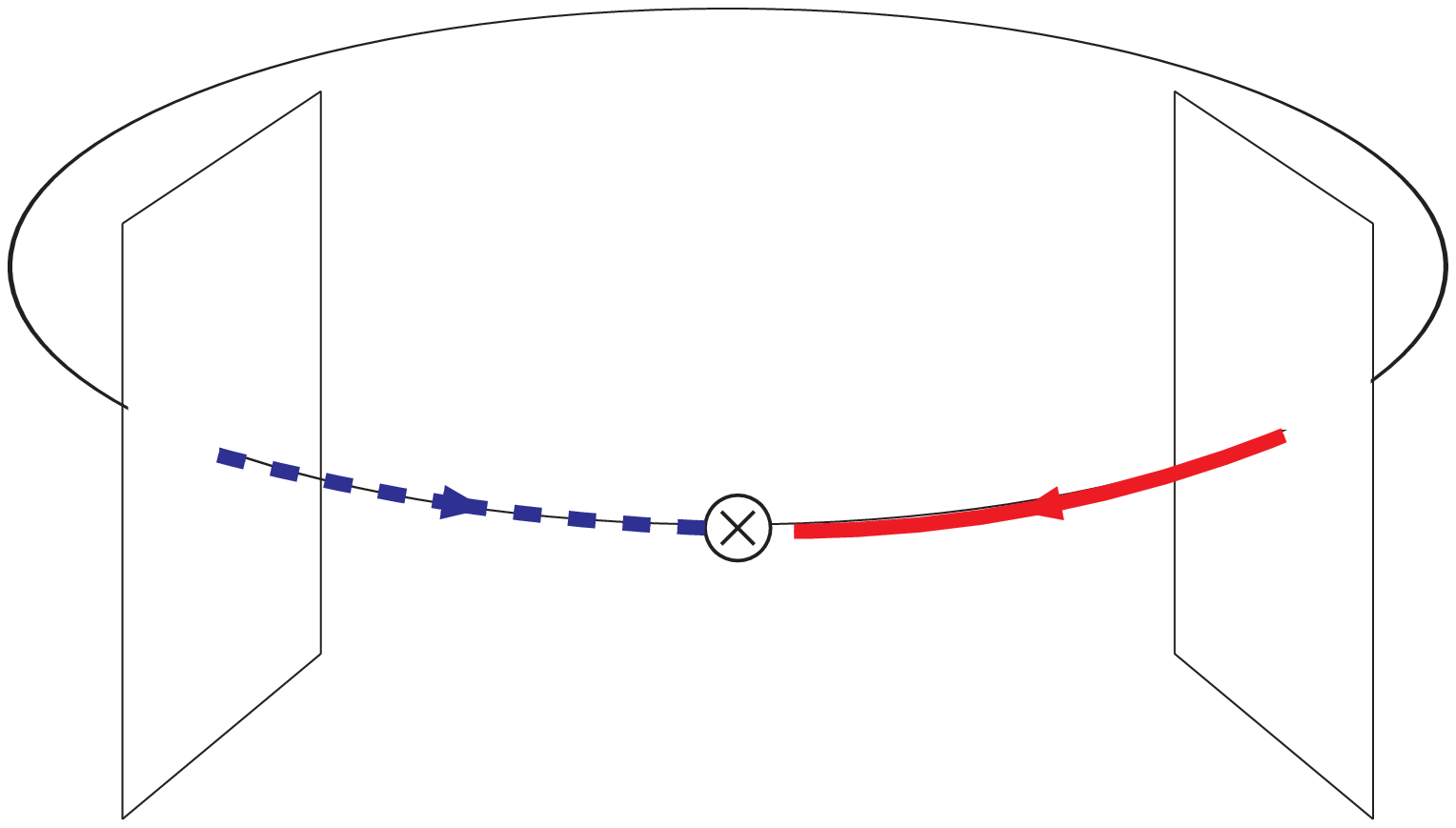} &
 \includegraphics*[width=0.4 \textwidth,clip=true]
  {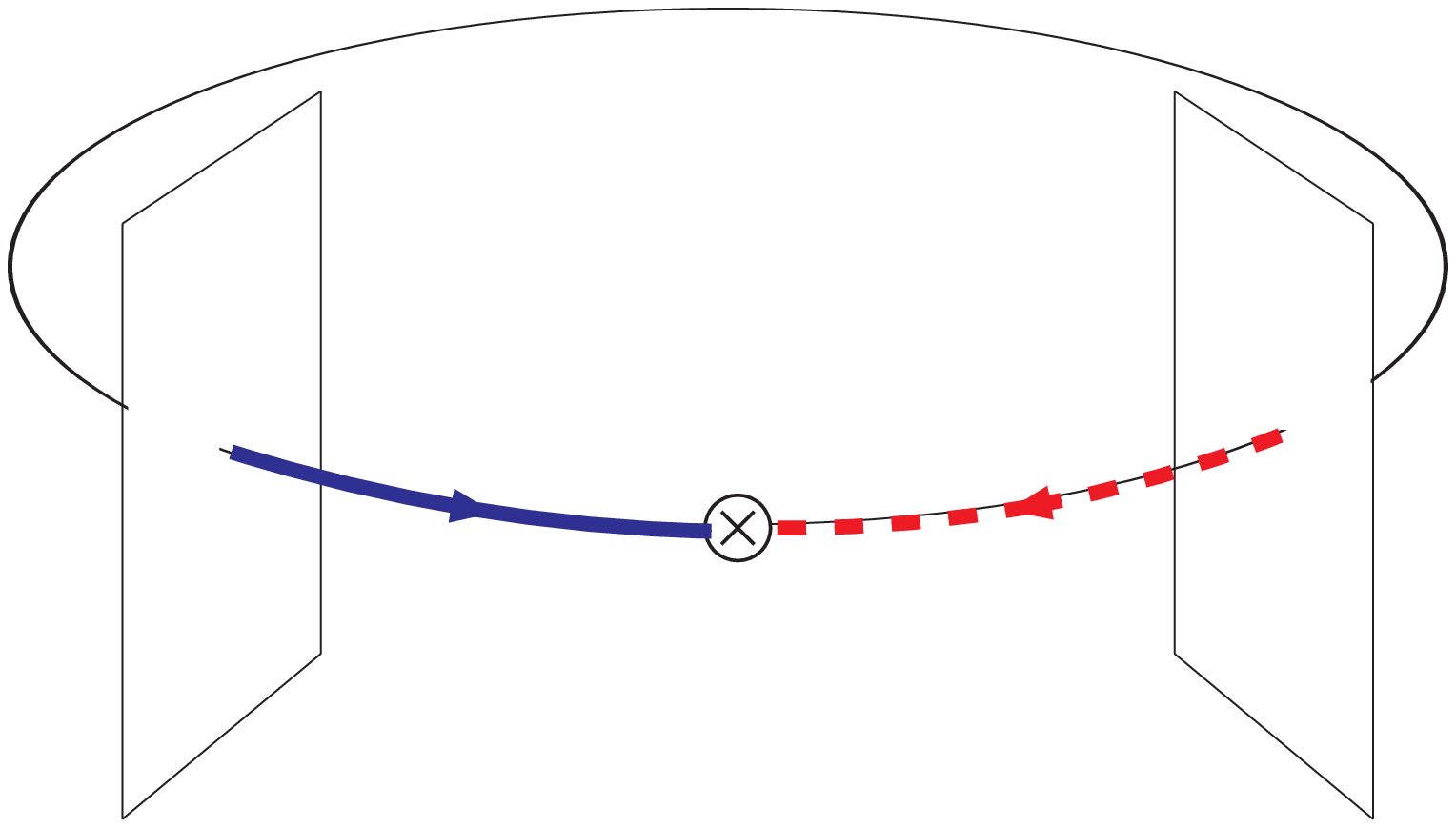}\\
 \includegraphics*[width=0.4 \textwidth,clip=true]
  {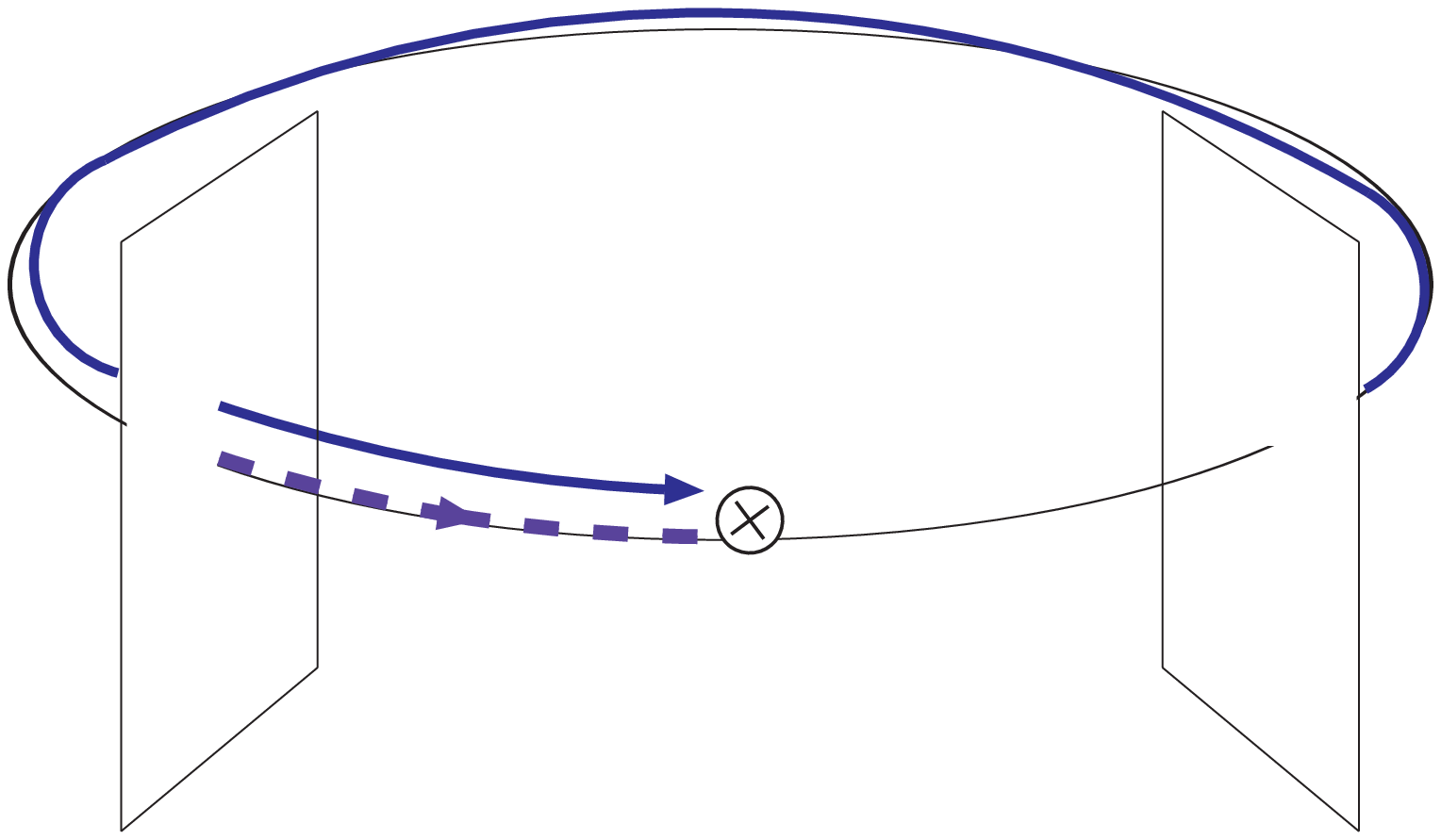} &
 \includegraphics*[width=0.4 \textwidth,clip=true]
  {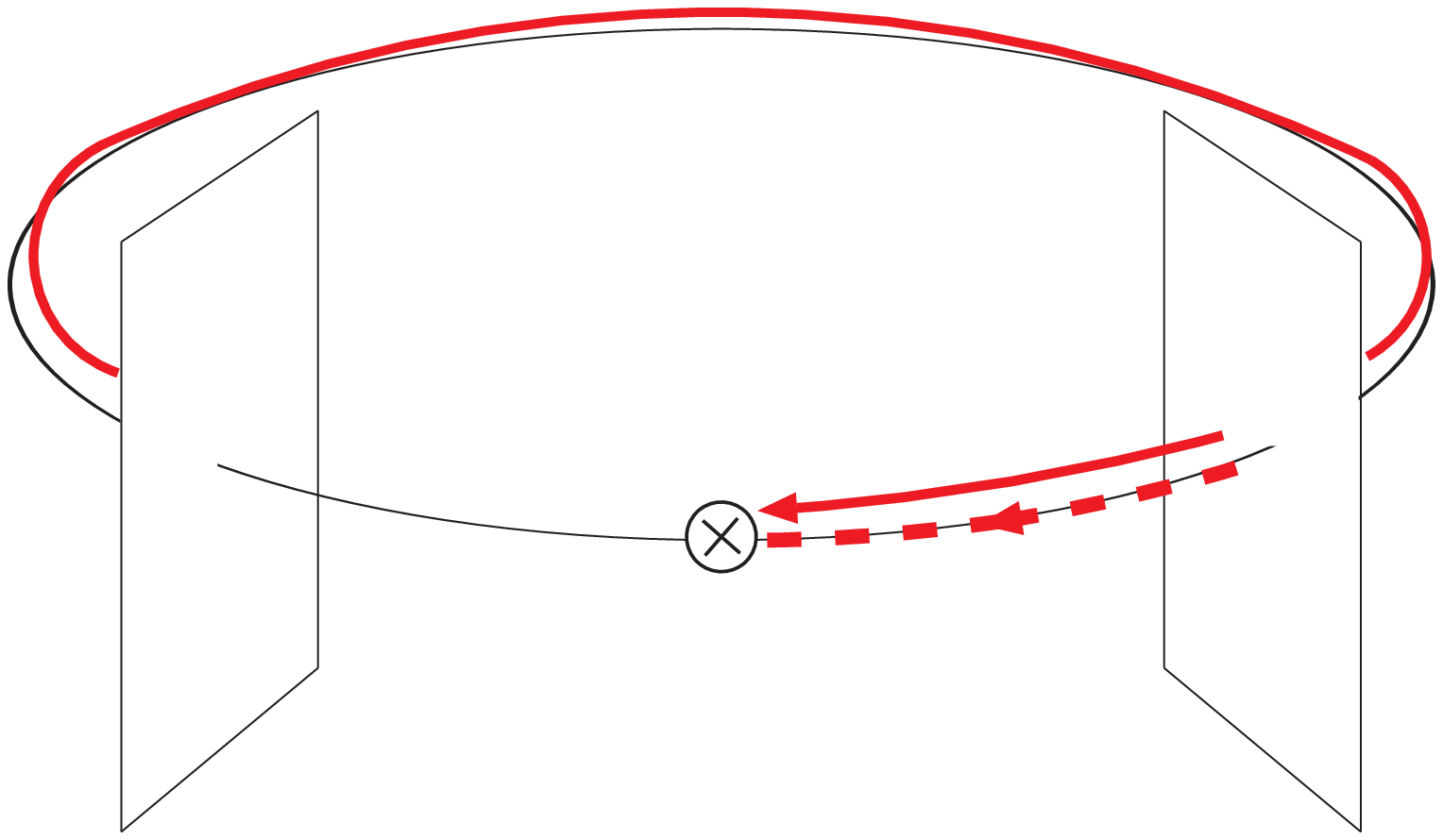}
 \end{tabular}
 \caption{Possible contaminations containing a wrapping or an excited
 state.
 The solid lines represent the kaon, and the dashed lines the first excited
 state.}
 \label{fig:contaminations}
\end{figure}

At large time separation, $|t_2-t|\gg 1$ and $|t-t_1|\gg 1$,
(\ref{eq:3pt-def}) is expected to behave as
\begin{eqnarray}
     \mbox{eq.~(\ref{eq:3pt-def})}
\rightarrow&&
     V_3\,c_0\,
     e^{-m_P(t_2-t_1)}\nonumber\\
 &+& V_3\,c_1\,
     e^{-m_P\,N_t-\Delta_P (t_2-t_1)/2}
     \cosh\bigg[(2\,m_P+\Delta_P)
      \left(t-\frac{t_2+t_1}{2}\right)\bigg]\no\\
 &+& V_3\,c_2\,
     e^{-(m_{P'}+m_P)\frac{t_2-t_1}{2}}
    \cosh\left[(m_{P'}-m_P)
               \bigg(t-\frac{t_2+t_1}{2}\bigg)\right]\nonumber\\
 &+& V_3\,c_3\,
 e^{-m_P N_t-(m_{P'}-m_P+\Delta'_P)(t_2-t_1)/2}
 \cosh\left[(m_{P'}+m_P+\Delta'_P)
            \left(t-\frac{t_2+t_1}{2}\right)
      \right].
 \label{eq:3pt-1}
\end{eqnarray}
The meaning of each term is explained in the following.
The first term in~(\ref{eq:3pt-1}) contains the hadron matrix
element relevant to the calculation of $B_K$, and 
$c_0=(Z^{\rm wall}_{A_4})^2\,
     \langle \bar P|O^{\rm lat}_{L_\mu L_\mu}\ |P \rangle/(2\,m_P)^2$.
The schematic diagram corresponding to this contribution is shown
in Fig.~\ref{fig:relevant}.
While this contribution does not depend on $t$, in the real data, as
shown in Fig.~\ref{fig:3pt}, the three-point function depends on $t$,
and the dependence is more pronounced for the lighter meson (left).
To describe this $t$-dependence, we incorporate three additional terms:
a contribution of a kaon wrapping around the lattice in the time
direction, an excited state contamination, and a mixture of a wrapping
kaon and an excited state.
The diagrams corresponding to these three contributions are depicted
in Fig.~\ref{fig:contaminations}.
The second term in~(\ref{eq:3pt-1}) represents a wrapping kaon
contribution [Fig.~\ref{fig:contaminations} (top)], and $c_1$ contains
the transition amplitude of a two-kaon state $|P, P\rangle$ to the
vacuum, $\langle 0|O^{\rm lat}_{L_\mu L_\mu}\ |P, P \rangle$.
$\Delta_P = E_{\rm total} - 2 m_P$ denotes the total energy difference
between an interacting and noninteracting two-kaon system, and is
extracted through the fit.
The third term describes the transition amplitude between the ground
state $|P\rangle$ and the first excited state  $|{\bar P}'\rangle$,
hence $c_2$ contains
$\langle {\bar P}'|\,O^{\rm lat}_{L_\mu L_\mu}\,|P\rangle$
[Fig.~\ref{fig:contaminations} (middle)].
The excited state mass $m_{P'}$ is extracted from the two-point function
with a point source and a point sink through a double cosh fit using the
fixed ground state mass.
Finally the fourth term represents the mixture contribution
[Fig.~\ref{fig:contaminations} (bottom)].

The set A and B are fitted simultaneously to obtain $c_0$ with $m_P$
fixed to the value extracted from the two-point function.
In the fit, two cases are examined, one taking the excited state
contaminations, {\it i.e.}, the $c_2$ and $c_3$ terms
in~(\ref{eq:3pt-1}), into account and the other not.
The $c_1$ term is always included.
In both cases, the fit range dependence is investigated.
For later use, the fit ranges $[\,t_{\rm min},\,t_{\rm max}]$ for the
set A and B are parametrized as 
\begin{eqnarray}
    [\,t_{\rm min},\,t_{\rm max}]
= \left\{\begin{array}{cc}
  {[\   8-dt,\ 8+dt]} &\ \ \ \mbox{for set A},\\
  {[\, 16-dt, 24+dt]} &\ \ \ \mbox{for set B}.
         \end{array}\right.
\end{eqnarray}
Then $8-dt$ corresponds to the time separation between an end of the fit
range and the nearest source operator.

\begin{figure}
 \centering
  \includegraphics*[width=0.6 \textwidth,clip=true]
  {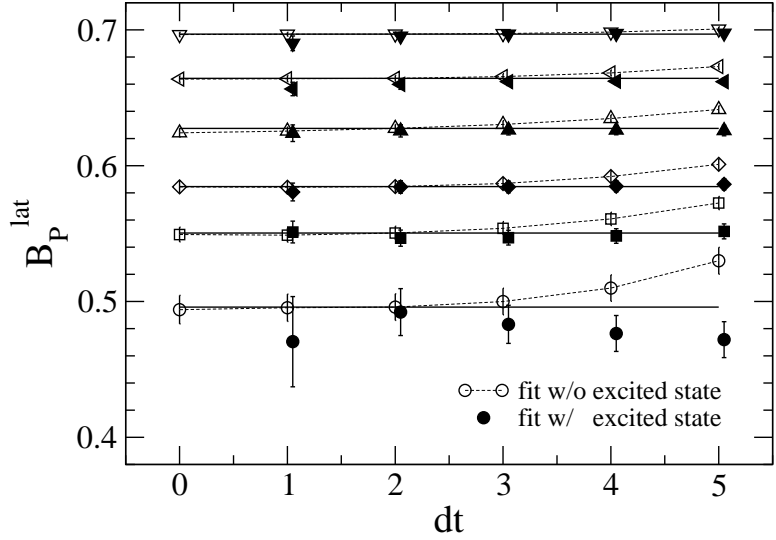}
 \caption{Fit range dependence of $B_P^{\rm lat}$.
 The open (filled) symbols are obtained by fitting the data
 to~(\ref{eq:3pt-1}) without (with) the $c_2$ and $c_3$ terms.
 The horizontal solid lines represent the value of the open symbol at
 $dt=2$.
 This analysis is performed only at the unquenched points
 ($m_{v1}=m_{v2}=m_{\rm sea}$).
 The quark mass increases from bottom to top.}
 \label{fig:bp-rangedep}
\end{figure}

We first perform an uncorrelated simultaneous fit without the $c_2$ and
$c_3$ terms.
The solid curves in Fig.~\ref{fig:3pt} represent the results for $dt=2$.
The numerical results of $c_0$, $c_1$ and $\Delta_P$ for $dt=2$ are
tabulated in
Tables~\ref{tab:mass-results-0.015}-\ref{tab:mass-results-0.050-Q-4}.
As seen from Fig.~\ref{fig:3pt}, the fit results are on the top of the
data within the error, and the $\chi^2$/dof values in the fit are
acceptable and in the range of [0.01, 0.2].
Although as seen from the table $\Delta_P$ is not determined very well
especially when the quark is light, the obtained values for $\Delta_P$
are reasonably consistent with those in~\cite{Yagi:lat07}, in which
$\Delta_P$ is calculated on the same configurations at
$m_{\rm sea}=m_{v1}=m_{v2}$ on the way to obtaining the $I$=2
$\pi$-$\pi$ scattering length, and is determined to a better precision.
The lattice $B$-parameter for a general pseudoscalar meson, which we
call $B_P^{\rm lat}$ in the following, is then obtained through
\begin{eqnarray}
     B_P^{\rm lat} 
 &=& \frac{3}{8}
     \left(\frac{2}{Z^{\rm wall}_{A_4}\,f_P}\right)^2
     \times
     \frac{(Z^{\rm wall}_{A_4})^2
           \langle \bar P|O^{\rm lat}_{L_\mu L_\mu}\ |P \rangle}
          {(2\,m_P)^2},
\end{eqnarray}
where the first and second factors are obtained from the two- and
three-point functions, respectively.
By varying $dt$, we obtain the $dt$-dependence of $B_P^{\rm lat}$ in
Fig.~\ref{fig:bp-rangedep} (open symbols), where the data with
$m_{\rm sea}$=$m_{v1}$=$m_{v2}$=$m_q$ are shown.
The results at $dt=2$ are indicated by the solid horizontal lines in
the figure.
It is seen that $B_P^{\rm lat}$ systematically increases at large $dt$
with $dt$ while it remains unchanged for $dt \le 2$.

\begin{figure}
 \centering
  \includegraphics*[width=0.6 \textwidth,clip=true]
  {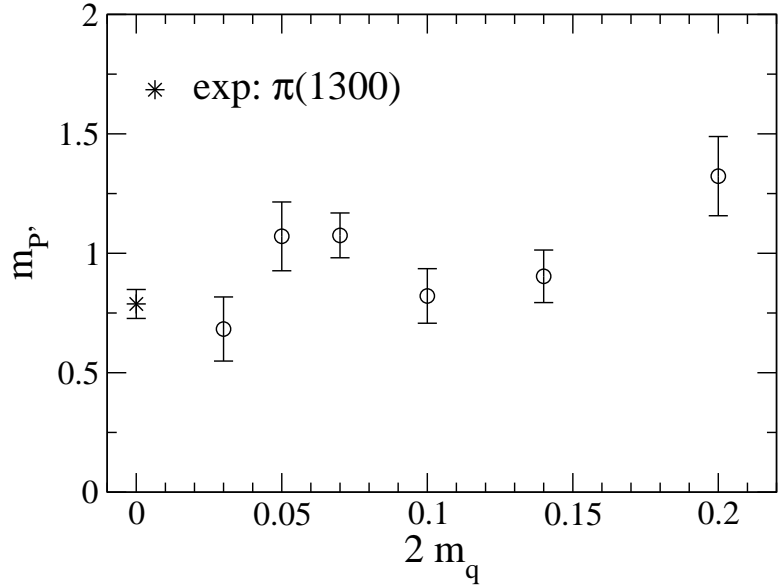}
 \caption{The quark mass dependence of the first excited state mass,
 $m_{P'}$ (circles) in the lattice unit.
 The experimental value of $\pi(1300)$ is also shown.}
 \label{fig:excited state mass}
\end{figure}

To make an analysis including the $c_2$ and $c_3$ terms, we use the
first excited state mass $m_{P'}$ extracted from the two-point function
with the point source and point sink.
Through a double cosh fit with a fixed ground state mass, we obtain
Fig.~\ref{fig:excited state mass}, which shows the first excited state
mass $m_{P'}$ as a function of $m_{\rm sea}$=$m_{v1}$=$m_{v2}$=$m_q$.
Although the statistical error is sizable, $m_{P'}$ reasonably
extrapolates to the experimentally measured value of the excited state
mass of $\pi$(1300).
In the following fits, $m_{P'}$ is fixed to the value thus extracted.
As seen from~(\ref{eq:3pt-1}), $\Delta_P'$ always appears in the
combination $m_{P'}+\Delta_P'$.
According to some trial fits in which $\Delta_P'$ is treated as a free
parameter, it turned out that $\Delta_P'$ is not determined well and its
size is similar to or even smaller than the statistical error of
$m_{P'}$.
Therefore, in the following analysis, whose purpose is to test the
stability against the fit range, we set $\Delta_P'$ to zero.
The simultaneous fit to~(\ref{eq:3pt-1}) including the $c_2$ and $c_3$
terms is carried out with varying $dt$ as before, and the resulting
$dt$-dependence of $B_P^{\rm lat}$ is shown in
Fig.~\ref{fig:bp-rangedep} (filled symbols).
It turns out that in this case $B_P^{\rm lat}$ does not depend on $dt$
and its value is consistent with the open symbol at $dt \le 2$ for all
the quark masses.

\begin{figure}
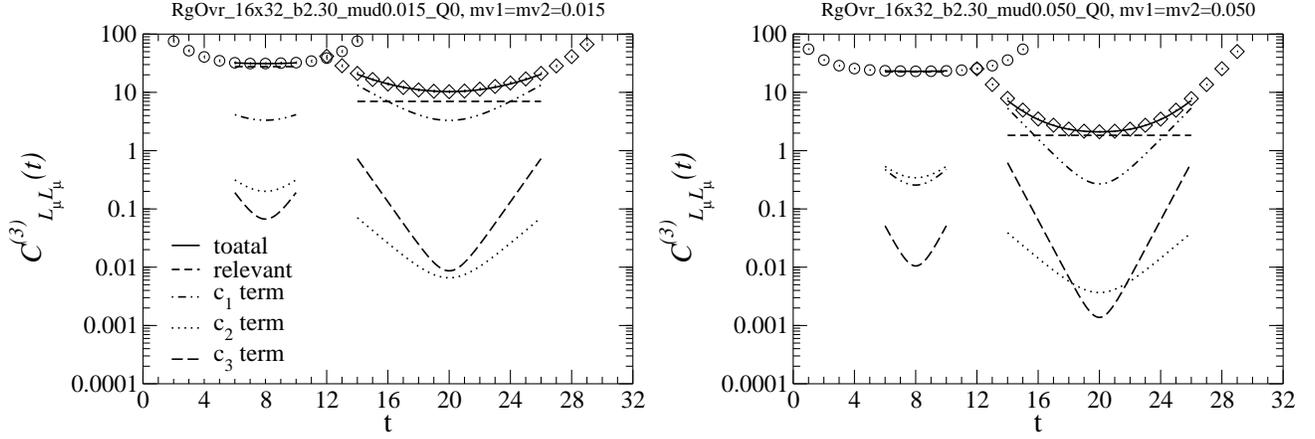

 \centering
 \begin{tabular}{cc}
  \includegraphics*[width=0.5 \textwidth,clip=true]
  {figs/3pt_LL_m0.015_001001-each.eps} &
  \includegraphics*[width=0.5 \textwidth,clip=true]
  {figs/3pt_LL_m0.050_004004-each.eps}
 \end{tabular}
 \caption{The size of each contribution.
 The plots for $m_{\rm sea}$=0.015 (left) and 0.050 (right) are shown
 as representatives.
 See the text for details.}
 \label{fig:3pt-each}
\end{figure}

To show the significance of each term in~(\ref{eq:3pt-1}), each
contribution is separately plotted with a logarithmic scale in
Fig.~\ref{fig:3pt-each}.
We find that the contributions from the $c_2$ and $c_3$ terms
in~(\ref{eq:3pt-1}) are always smaller than the others.
An exception is the set A at $m_{\rm sea}$=0.050 (right), in
which the $c_2$ term is as large as the $c_1$ term.
But in this case the size of these contributions is only a few \% of the
relevant term to extract $B_P^{\rm lat}$.
Another possible contamination, which has not been discussed so far, is
the one containing the transition amplitude between the first excited
states, $\langle {\bar P}'|\,O^{\rm lat}_{L_\mu L_\mu}\,|P'\rangle$.
Using $m_{P'}$ obtained above, its effect to $B_P^{\rm lat}$ is
estimated to be less than 0.03 \%, and so is neglected.

From the above observations, in the following analysis we use
$B_P^{\rm lat}$ obtained at $dt=2$ without the $c_2$ and $c_3$ terms.

\subsection{Nonperturbative matching}
\label{subsec:npr}

\begin{figure}
 \centering
  \includegraphics*[width=0.6 \textwidth,clip=true]
  {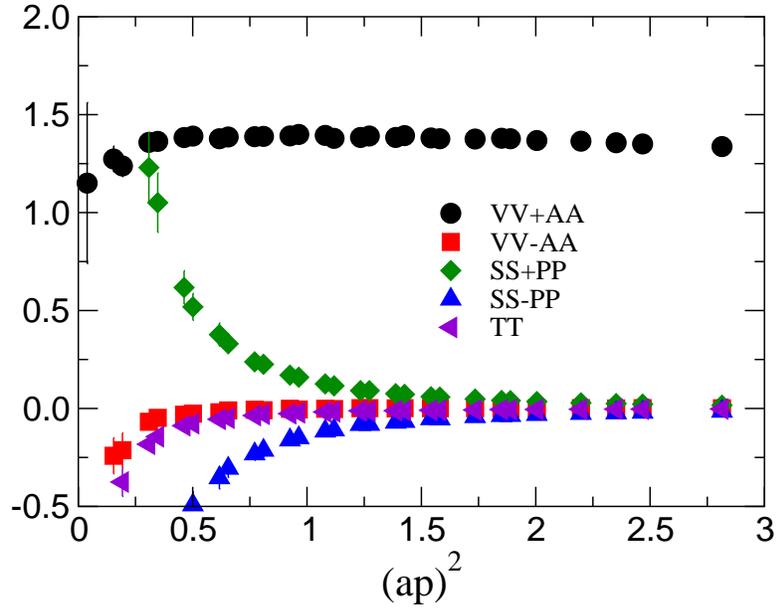}
 \caption{Lattice momentum dependence of each component of the
 five-point vertex function for $m_q=0.015$.}
 \label{fig:npr mixing}
\end{figure}

We adopt the RI/MOM scheme~\cite{Martinelli:1994ty} to calculate the
matching factor.
We follow the standard method, which is briefly described in the
following.
In this subsection, we consider the renormalization of the operator of
the chiral structure
$VV+AA$,
$(\bar q_1 \gamma_\mu\,q_2'\ \bar q_1 \gamma_\mu\,q_2')+
 (\bar q_1 \gamma_\mu\gamma_5\,q_2'\ \bar q_1\gamma_\mu\gamma_5\,q_2')$,
rather than $O_{L_\mu L_\mu}$ defined in (\ref{eq:four-quark-op}),
since in the presence of chiral symmetry the matching factors for these
two operators are equivalent.
Fixing the gauge to the Landau gauge, we first calculate the five-point
vertex function (or equivalently the amputated five-point function) for
the $VV+AA$ operator, where four external off-shell momenta are set to a
common value.
By applying a proper spin-color projection, the vertex function is
decomposed into five different structures,
$(VV\pm AA),\ (SS\pm PP),\ TT$, where $V,A,S,P,T$ denote vector,
axial-vector, scalar, pseudoscalar, and tensor bilinears, respectively.
In Fig.~\ref{fig:npr mixing}, the lattice momentum dependence of the
multiplicative part (circles) and the mixing part (other symbols) are
shown.
As is guaranteed by the exact chiral symmetry, the nonmultiplicative
contributions are strongly suppressed and vanish asymptotically as
momentum becomes large.
By imposing the renormalization condition that the $VV+AA$ component of
the renormalized vertex function be equal to the tree-level value, we
obtain $Z_q^{-2}\,Z_{VV+AA}$ with $Z_q$ the quark wave function
renormalization.
We also calculate the vertex function for the axial-vector current with
the same momentum configurations to obtain $Z_q^{-1}\,Z_{A}$.
Taking a ratio of the multiplicative part of the five-point vertex
function to a square of the vertex function for the axial-vector
current, we obtain $Z_{B_K}^{\rm RI/MOM}=Z_{VV+AA}/Z_{A}^2$ at each
quark mass and each momentum.

The momentum $a p_\mu$ is defined by $a p_\mu = 2\pi n_\mu/L_\mu$,
where $L_\mu$ is the number of total lattice sites in the $\mu$
direction and $n_\mu$ is an integer.
While $n_\mu$ can take the value in the range of
$[-(L_\mu/2)+1, (L_\mu/2)]$, in order to avoid the large discretization
error we restrict the range to that satisfying $a p_\mu < 1$.
Namely, $n_\mu$ can only take
$n_i = \{-2, -1, 0, 1, 2\}$ for $i=x, y, z$ and
$n_t = \{-5, -4, -3, -2, -1, 0, 1, 2, 3, 4, 5\}$.
Then the maximum value for $(a p)^2$ used in the following analysis is
about 2.81.

\begin{figure}
 \centering
  \includegraphics*[width=0.6 \textwidth,clip=true]
  {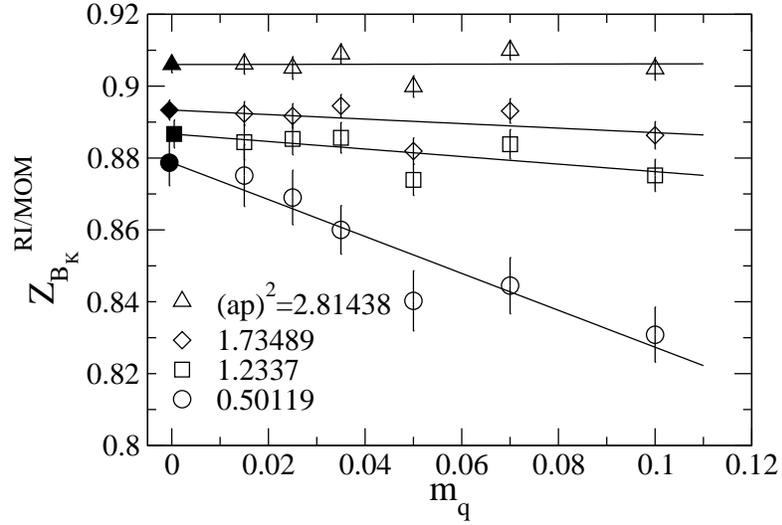}
 \caption{Chiral extrapolation of $Z_{B_K}^{\rm RI/MOM}$ at several
 representative momenta.}
 \label{fig:npr chiral extrp}
\end{figure}

\begin{figure}
 \centering
  \includegraphics*[width=0.6 \textwidth,clip=true]
  {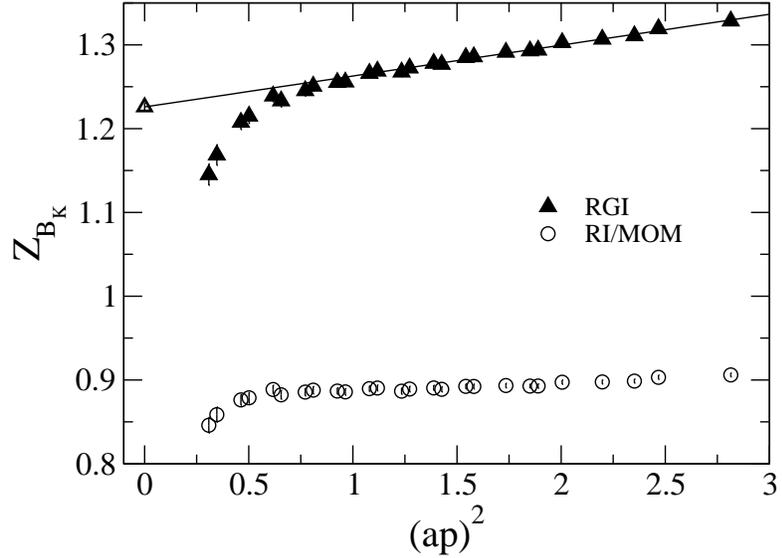}
 \caption{$Z_{B_K}^{\rm RI/MOM}$ in the chiral limit and
 $Z_{B_K}^{\rm RGI}$ as a function of the external state momentum
 squared.}
 \label{fig:npr zrgi}
\end{figure}

The chiral extrapolation of $Z_{B_K}^{\rm RI/MOM}$ is made
linearly in the quark mass as shown in Fig.~\ref{fig:npr chiral extrp}.
A clear dependence on quark mass is observed at relatively small
$(ap)^2$ while the dependence vanishes at larger $(ap)^2$.
After the chiral limit at each lattice momentum, we finally obtain
$Z_{B_K}^{\rm RI/MOM}$ shown in Fig.~\ref{fig:npr zrgi} (open
circles).
In~\cite{Aoki:2007xm}, possible nonperturbative contaminations are
discussed in detail.
With the momentum setup commonly used in the RI/MOM scheme, some
nonperturbative effects may be enhanced in the small momentum region.
They are responsible for the linear dependence on the quark mass in this
region.
To avoid such contaminations, in the following analysis we restrict the
data point to those of $(ap)^2>1.2337$, where the linear slope is
consistent with zero within 2 standard deviations.

From $Z_{B_K}^{\rm RI/MOM}(p^2)$ the renormalization group invariant
(RGI) factor $Z_{B_K}^{\rm RGI}$ is obtained by
\begin{eqnarray}
&&  Z_{B_K}^{\rm RGI}
=w^{-1}_{\rm RI/MOM}(\mu^2)\,Z_{B_K}^{\rm RI/MOM}(\mu^2),
\end{eqnarray}
where 
\begin{eqnarray}
 w_{\rm RI/MOM}(\mu^2)
=\left(\alpha_s(\mu^2)\right)^{\frac{\gamma_0}{2\beta_0}}
 \left[1-\frac{\alpha_s(\mu^2)}{4\pi}J_{\rm RI/MOM}\right],
 \label{eq:running-factor}
\end{eqnarray}
is the running factor due to the anomalous dimension at the
next-to-next-to-leading order.
$\gamma_0=4$ and
\begin{eqnarray}
 J_{\rm RI/MOM}
=\frac{-(23931-2862 N_f+128 N_f^2)}{6\,(33-2 N_f)^2}+1+8\ln 2,
\label{eq:J_rimom}
\end{eqnarray}
which is calculated in~\cite{Ciuchini:1997bw}.
We use the running coupling constant at the same order given by
\begin{eqnarray}
 \alpha_s(\mu^2)
=\frac{4\pi}{\beta_0 L}
 \left[1-\frac{\beta_1}{\beta_0^2}\frac{\ln L}{L}
 \right],
\end{eqnarray}
with $\beta_0=\frac{33-2 N_f}{3}$, $\beta_1=102-10 N_f-\frac{8}{3}N_f$
and $L=\ln(\mu^2/\Lambda_{\rm QCD}^2)$.
In~\cite{Della Morte:2004bc}, $\Lambda_{\rm QCD}$ for $N_f=2$ in the
$\overline{\rm MS}$ scheme is calculated to be 245(16)(16) MeV assuming
$r_0$=0.5 fm.
By summing up the two errors and converting to $r_0$=0.49 fm, we obtain
$\Lambda_{\rm QCD}=250(33)$ MeV which we will use in the following
analysis.

The resulting $Z_{B_K}^{\rm RGI}$ is shown in Fig.~\ref{fig:npr zrgi}
(filled triangles).
$Z_{B_K}^{\rm RGI}$ must be independent of the renormalization scale up
to neglected higher order effects.
The remaining scale dependence due to the neglected higher order effects
is estimated as follows.
In our momentum region, the factor $1/w_{\rm RI/MOM}(\mu^2)$ is well
approximated by a linear function of $(ap)^2$ as
$w_0\,[1+w_1\,(ap)^2]$.
The slope $w_1$ determined with two points $(ap)^2$=1.2337 and 2.81438
is 0.017 or 0.022, with and without the $O(\alpha_s)$ term
in~(\ref{eq:running-factor}), respectively.
From the difference of $w_1$ between these two cases, we deduce that the
$O(\alpha_s^2)$ correction affects $w_1$ by less than 0.005.
On the other hand, fitting the data to a linear function of $(ap)^2$, we
obtain $Z_{B_K}^{\rm RGI}=1.226(5)\times[1+0.030(2)\times(ap)^2)]$.
Therefore, we conclude that the remaining $(ap)^2$ dependence is
dominated by the $O(a^2p^2)$ discretization effects, which are removed
by a linear extrapolation to $(ap)^2=0$ as shown in
Fig.~\ref{fig:npr zrgi}. 
We observe a clear nonzero slope in $(ap)^2$ for $Z_{B_K}^{\rm RGI}$,
but the coefficient is sufficiently small to rely on the extrapolation.

To follow the standard convention, the matching factor is converted to
that in the $\overline{\rm MS}$ scheme using
\begin{eqnarray}
 Z_{B_K}^{\rm \overline{MS}}(\mu^2)
=w_{\rm \overline{MS}}(\mu^2)\,Z_{B_K}^{\rm RGI},
\end{eqnarray}
where $w_{\rm \overline{MS}}(\mu^2)$ is obtained from
(\ref{eq:running-factor}) by replacing $J_{\rm RI/MOM}$
with~\cite{Ciuchini:1997bw}
\begin{eqnarray}
 J_{\rm \overline{MS}}
=\frac{-(23931-2862 N_f+128 N_f^2)}{6\,(33-2 N_f)^2}+\frac{17}{3}.
\end{eqnarray}
We obtain $w_{\rm \overline{MS}}(\mu^2)=0.7086$ at $\mu=$2 GeV.

In the whole procedure, the largest uncertainty comes from the
perturbative running.
The matching factor at a given scale $\mu$ in the RI/MOM scheme is
obtained nonperturbatively, but its conversion to other schemes is not.
The systematic uncertainty is then estimated by the size of the
correction of the highest order included (next-to-leading order in this
work), which is 0.071 for $Z_{B_K}^{\rm RGI}$ and 0.020 for
$Z_{B_K}^{\overline{\rm MS}}(2\ {\rm GeV})$.
The uncertainty of $\Lambda_{\rm QCD}$ gives $\pm$0.013 for
$Z_{B_K}^{\rm RGI}$ and $\pm$0.0002 for
$Z_{B_K}^{\overline{\rm MS}}(2\ {\rm GeV})$.
These systematic errors are added in quadrature.
Finally we obtain
\begin{eqnarray}
Z_{B_K}^{\rm RGI}=1.224(5)(72),\ \ \
Z_{B_K}^{\overline{\rm MS}}(2\ {\rm GeV})=0.867(3)(20),
\end{eqnarray}
where the first error is statistical and the second one is systematic.
The results for the $B$-parameter in the $\overline{\rm MS}$ scheme at
$\mu$=2 GeV, $B_P^{\overline{\rm MS}}(2 {\rm GeV})$, are given in
Tables~\ref{tab:mass-results-0.015}-\ref{tab:mass-results-0.050-Q-4}.

\section{Test with NLO PQChPT}
\label{sec:test-PQChPT}

Before extrapolating the lattice data to the physical kaon mass
($m_{v1}=m_{\rm sea}\rightarrow m_{ud}^{\rm phys},\
  m_{v2}\rightarrow m_s^{\rm phys}$),
we test whether the quark mass dependence of $B_P$ is consistent with
the NLO PQChPT
prediction.
In other words we try to identify the quark mass region, where the
lattice results are well described by ChPT.
We restrict the data points used to those which satisfy
$m_{\rm sea}\le m_{\rm valence}$ for the reason described below.

In~\cite{Becirevic:2003wk}, the finite volume effect (FVE) to $B_K$ is
studied to NLO in the framework of finite volume PQChPT, where the
``kaon'' consists of a light quark with mass $m_{v1}$ and the physical
strange quark mass with $m_{v2}$ fixed to $m_s^{\rm phys}$.
Its numerical results indicate that the FVE is more profound as $m_{v1}$
vanishes while $m_{\rm sea}$ is fixed.
This can be deduced from the NLO PQChPT formula for $B_K$ in the infinite
volume alone, because it contains the term proportional to
$m_{\rm sea}\ln (m_{v1}/m_{v2})$ and the loop integral leading to this
term is expected to be sensitive to the infrared cutoff, or equivalently
to the size of the spatial volume.
Then the chiral expansion with $m_{v1}\ll m_{\rm sea}$ becomes unlikely
to converge quickly and hence less reliable.
Quantitatively, the estimate of the FVE to $B_K$
in~\cite{Becirevic:2003wk} gives about 3 \% for
$m_{v1} \sim m_s^{\rm phys}/5$ and
$m_{v2}=m_{\rm sea} \sim m_s^{\rm phys}$ and so appears to be under
control.
However it is pointed out in~\cite{Colangelo:2005gd} (and cautioned
in~\cite{Becirevic:2003wk}) that at the NLO the FVE could be
significantly underestimated for $m_{\pi}$ and $f_\pi$.
For example, the NLO estimate of the FVE to $f_\pi$ gives about 2 \%
correction at our lightest unquenched point while the inclusion of NNLO
gives 4\%--5 \%.
It should be noted that this study is made at the unquenched points
($m_{\rm sea}=m_{v1}=m_{v2}$).
Therefore it could be worse when $m_{v1}\ll m_{\rm sea}\sim m_{v2}$
because of the above reason.
Motivated by these observations, in the following analysis we include
the data points only when $m_{\rm sea}\le m_{\rm valence}$.

In this test, we focus on the data points consisting of degenerate
quarks ($m_{v1}=m_{v2}$) for simplicity.
The NLO PQChPT prediction for $B_P$ with degenerate valence quarks
is~\cite{Golterman:1997st,Becirevic:2003wk},
\begin{eqnarray}
     B_P
 &=& B_P^\chi\Bigg[\,
       1- \frac{6\,m_P^2}{(4\pi f)^2}
          \,\ln\left(\frac{m_P^2}{\mu^2}\right)\Bigg]
     + (b_1-b_3)\,m_P^2 + b_2\,m_{ss}^2,
 \label{NLOform}
\end{eqnarray}
where $m_{ss}$ is the pseudoscalar meson mass with
$m_{v1}=m_{v2}=m_{\rm sea}$ and $B^\chi_P$, $f$, $(b_1-b_3)$, and $b_2$
are free parameters.
$f$ is the tree-level pion decay constant in the $f_\pi\sim$130 MeV
normalization, and is the only parameter which controls nonlinear
dependence of $B_P$ on the pseudoscalar meson mass squared.
In the fit, $\mu$ is set to 1 GeV.
The numerical data are fitted to (\ref{NLOform}) with a varying fit
range.

\begin{figure}
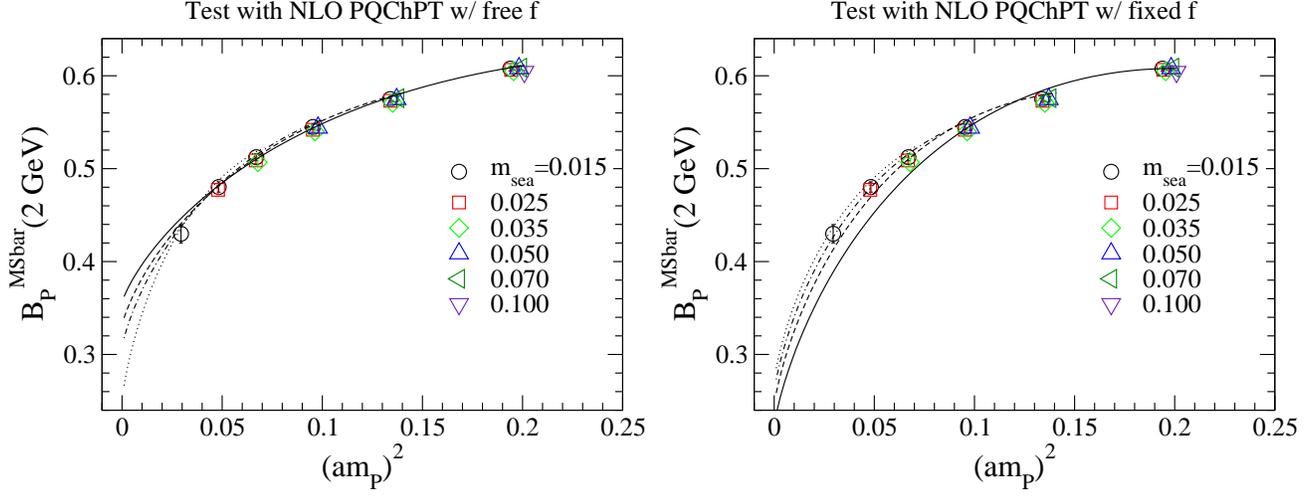

 \centering
  \includegraphics*[width=0.5 \textwidth,clip=true]
  {figs/bp-nlo-mv1=mv2_ffree-limit.eps}~
  \includegraphics*[width=0.5 \textwidth,clip=true]
  {figs/bp-nlo-mv1=mv2_ffix-limit.eps}\\[-1ex]
 \caption{$B_P^{\overline{\rm MS}}$(2\,GeV) as a function of valence
 pion mass squared.
 The data are fit to the NLO PQChPT formula with an unfixed $f$ (left)
 and the fixed $f=0.0659$ (right).
 The different symbols denote different sea quark masses.
 Several curves are obtained with different fit ranges and show the ones
 extrapolated to $m_{\rm sea}=0$.}
 \label{fig:bp-valdep-fit2-limit}
\end{figure}

The fit results are summarized in Table~\ref{tab:nlo-test} (top) and
shown in Fig.~\ref{fig:bp-valdep-fit2-limit} (left).
While all fit ranges tested give acceptable $\chi^2/$dof, $f$
monotonically increases as the fit range is extended.
$f$'s obtained from the two narrowest ranges are consistent with each
other within 1 standard deviation, and also consistent with a naive
expectation $100\sim 130$ MeV.
We also attempt another fit fixing $f$ to 0.0659, which corresponds to
110 MeV obtained in our separate calculation~\cite{Noaki:2007es}.
The numerical results and plot are given in Table~\ref{tab:nlo-test}
(bottom) and in Fig.~\ref{fig:bp-valdep-fit2-limit} (right),
respectively.
The $\chi^2$/dof is acceptable for the two narrowest fit ranges, and the
results for other fit parameters are consistent with the values obtained
without fixing $f$.
From these observations, we conclude that the data for $m_q\le 0.050$
are inside the NLO regime while the data at around the strange quark
mass are not.

\section{Extraction of $B_K$}
\label{sec:extraction of BK}

Since the NLO PQChPT formula describes the data only up to about a half
of the physical strange quark mass, to extract $B_K$ at physical quark
masses from the data, it is necessary to modify the NLO PQChPT formula.
We incorporate an analytic term into the original NLO PQChPT
formula~\cite{Golterman:1997st,Becirevic:2003wk,Aoki:2004ht} as
\begin{eqnarray}
     B_{12}
 &=& B_{12}^\chi\Bigg[
       1- \frac{2}{(4\pi f)^2}
       \Bigg\{ m_{ss}^2 + m_{11}^2 -
       \frac{3\,m_{12}^4+m_{11}^4}{2\,m_{12}^2}
        + m_{12}^2\left(     \ln\left(\frac{m_{12}^2}{\mu^2}\right)
                     + 2\,\ln\left(\frac{m_{22}^2}{\mu^2}\right)
               \right)
\nonumber\\&&\hspace*{18ex}
       - \frac{1}{2}\left(
         \frac{m_{ss}^2(m_{12}^2+m_{11}^2)}{2\,m_{12}^2} 
                          + \frac{m_{11}^2
                          (m_{ss}^2-m_{11}^2)}
                          {m_{12}^2-m_{11}^2}
                    \right)
       \ln\left(\frac{m_{22}^2}{m_{11}^2}\right)
       \Bigg\}\Bigg]
\nonumber\\&&\hspace*{0ex}
     + b_1\,m_{12}^2
     + b_3\,m_{11}^2\left(-2+\frac{m_{11}^2}{m_{12}^2} \right)
     + b_2\,m_{ss}^2
     + d_1\,(m_{12}^2)^2,
 \label{eq:NLOform-nondege}
\end{eqnarray}
where $m_{ij}$ is the pseudoscalar meson mass consisting of valence quarks
$i$ and $j$.
In the limit of $m_{v1}=m_{v2}$, the above formula without the last
term reduces to (\ref{NLOform}).
The formula (\ref{eq:NLOform-nondege}) without the last term is the
NLO PQChPT
prediction for nondegenerate valence
quarks~\cite{Golterman:1997st,Becirevic:2003wk,Aoki:2004ht}, and the
last term is added to interpolate the data in the heavy valence quark
mass region.
Since the modification is used to interpolate the existing data in the
heavier valance quark mass region, its precise form is irrelevant to the
final result.
Actually we have confirmed that introducing the other term 
$d_2 (m_{11}^2m_{22}^2)$ into the formula changes the final result by
0.2 \% at most.

\begin{figure}
 \centering
  \includegraphics*[width=0.6 \textwidth,clip=true]
  {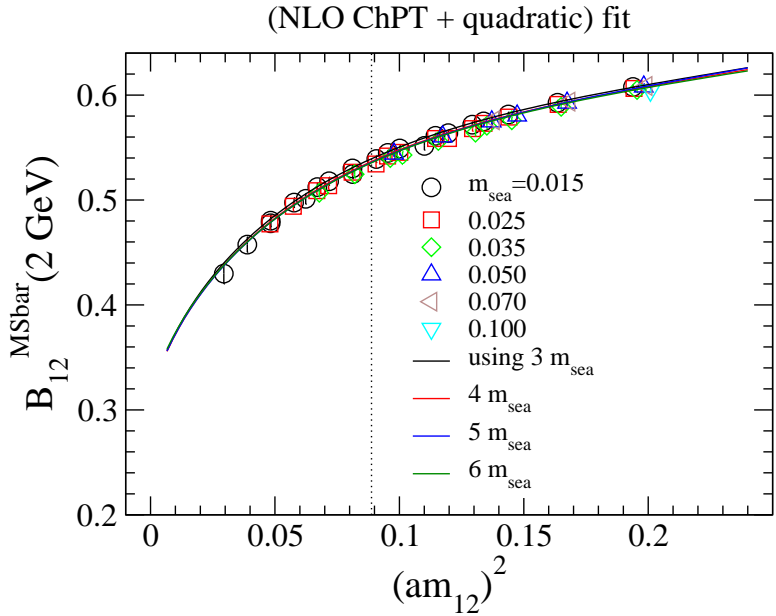}
 \caption{
 $m_{12}^2$ dependence of $B_{12}^{\overline{MS}}$(2\,GeV).
 The different symbols correspond to different $m_{\rm sea}$.
 The solid curves represent $B_P$ with $m_{\rm sea}$ and $m_{v1}$
 extrapolated to $m^{\rm phys}_{ud}$.
 The vertical line indicates the location of physical $m_K$.}
 \label{fig:bp-nnlo-nondege-limit}
\end{figure}

The fit is performed with four data sets, each of which includes the
data of three, four, five or six lightest sea quarks.
All the data satisfying $m_{\rm sea} \le \min(m_{v1}, m_{v2})$ are
included in the fit.
Numerical results and the plots are given in Table~\ref{tab:nnlo} and
Fig.~\ref{fig:bp-nnlo-nondege-limit}, respectively.
Solid curves in Fig.~\ref{fig:bp-nnlo-nondege-limit} represent the fit
result extrapolated to the point where one of the valence quark mass and
the sea quark mass ($m_{\rm sea}$) are equal to the physical $u, d$ mass
($m^{\rm phys}_{ud}$) while the other valence quark mass is free.
Therefore the lines show the strange (or heavier valence) quark mass
dependence of the $B$-parameter.
Since the four curves obtained with different data sets are
indistinguishable from each other and all the data used in the fit are
on top of each other, we may conclude that the difference between
degenerate and nondegenerate data is negligible for $B_P$.
As seen from Table~\ref{tab:nnlo}, $B_P^\chi$'s, $b_2$'s and $f$'s for
all the fit ranges are reasonably consistent with those obtained for the
two narrowest fit ranges with unfixed $f$ in the NLO test shown in
Table~\ref{tab:nlo-test} (top).
Interpolating to physical $m_K^2$, we obtain
$B_K^{\overline{\rm MS}}(2\ {\rm GeV})$, listed in Table~\ref{tab:nnlo}.
We take $B_K^{\overline{\rm MS}}(2\ {\rm GeV})=0.537(4)$, which is the
result with the fit range [0.015, 0.050], as the central value.
The difference from the others ($\pm 0.002$) is ignored as it is
much smaller than other systematic errors.

\section{Systematic errors}
\label{sec:systematic errors}

\subsection{Finite volume effects}

With our lattice size and the lightest pion mass, $m_\pi L$ is slightly
smaller than 3, for which one expects sizable finite volume effects.
One of such effects, which is special in the partially quenched theory
and becomes significant when $m_{v1},m_{v2}<m_{\rm sea}$, has been
eliminated by omitting a potentially dangerous data set.
If we apply the estimate based on the finite volume NLO PQChPT
analysis~\cite{Becirevic:2003wk} to our lattice setup with $L$=2 fm, the
finite size effect is estimated to be less than 1 \% over all the data points
we have included in the fit.
However, as mentioned before, the NNLO analysis revealed that the NLO
analysis could significantly underestimate the effect for $m_{\pi}$ and
$f_\pi$~\cite{Colangelo:2005gd}.
Unfortunately, the NNLO calculation of $B_K$ is not available.
Therefore, we add 5 \% uncertainty, which is the NNLO estimate on
$f_\pi$ at our lightest quark mass~\cite{Noaki:2007es}, as a
conservative upper bound of the finite volume effect.

\subsection{Fixing topology}

\begin{figure}
 \centering
  \includegraphics*[width=0.6 \textwidth,clip=true]
  {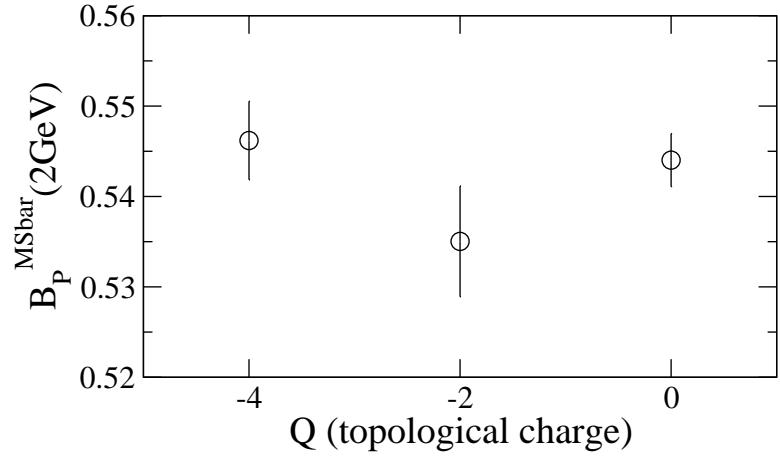}
 \caption{
 Comparison of $B_P^{\overline{MS}}$(2\,GeV) at $m_{\rm sea}$=0.05 with
 three different topological charges $Q=0,\ -2,\ -4$.
 }
 \label{fig:bp-nnlo-nondege-limit-topo}
\end{figure}

In our calculation there is a finite volume effect of a different
origin, {\it i.e.}, the fixed topological charge.
This effect is suppressed for large volumes as $1/V$, and is calculable
provided that the topological susceptibility and the $\theta$-dependence
of the quantity of interest are known~\cite{Brower:2003yx,Aoki:2007ka}.
The topological susceptibility is calculated on the same set of
lattices~\cite{Aoki:2007pw}.
Within the framework of ChPT, it can be shown that the most significant
$\theta$-dependence of the physical quantities is that of pion mass and
other quantities are affected through it.
We estimate the size of the effect on $B_P$ as
\begin{eqnarray}
 \sim B^\chi_P \frac{m_P^2}{(4\pi\,f)^2}\frac{1}{\langle Q^2 \rangle}
      \left(1-\frac{Q^2}{\langle Q^2 \rangle}\right),
\label{eq:assumption}
\end{eqnarray}
which appears at the next-to-leading order of ChPT.
Here, $\langle Q^2 \rangle=\chi_t V_4\sim 10$ at
$m_q=0.05$~\cite{Aoki:2007pw} and $V_4$ is the four-dimensional volume.
At this sea quark mass, the correction to the $Q=0$ result is estimated
to be 1.4\%, and the difference between $Q=0$ and $-2$ $(-4)$ to be
0.6\% (2.4\%).
To see whether this expected difference is seen or not, three results of
$B_P$ at $m_q$=0.05 in the $Q$=0, $-2$, $-4$ sectors are compared in
Fig.~\ref{fig:bp-nnlo-nondege-limit-topo}.
Since the size of the statistical error for the measured $B_P$ is about
1\% or so as shown in the figure, we do not expect clear systematic
$Q$-dependence of $B_P$, that is confirmed by the numerical data.
From this observation, we can safely assume that~(\ref{eq:assumption})
gives a reasonable estimate.
We quote 1.4\ \% as an estimate for the systematic error due to fixing
the topological charge.
A more complete analysis is available for $f_\pi$~\cite{Noaki:2007es}, for
which the effect of fixing the topological charge is estimated to be
about 1\ \% or less depending on the quark mass.

\subsection{Other systematic errors}

In addition to the above errors, we estimated 2 \% uncertainty in the
determination of $Z_{B_K}^{\overline{\rm MS}}(2\ {\rm GeV})$. 
Since the calculation is made only at one lattice spacing, a reliable
estimate of the scaling violation is difficult.
There is, however, an indication that the overlap fermion formulation
has relatively small scaling violation within the quenched calculation
of $B_K$~\cite{Babich:2006bh}, where no significant dependence on the
lattice spacing was observed between $1/a\sim$ 2.2~GeV and 1.5~GeV. 
Even at a fixed lattice spacing, the use of a different input to fix
the lattice spacing could lead to an appreciable change of $B_K$
because $B_K$ has a significant dependence on the squared meson mass
$(a m_{12})^2$ as seen in Fig.~\ref{fig:bp-nnlo-nondege-limit}.
It turns out that if one changes $1/a$ by $\pm$ 5 \% for instance, $B_K$
is changed by $\mp$ 5 \%.

This calculation is made with two flavors of dynamical quarks, and the
strange quark is quenched.
In~\cite{Aoki:2004ht} and \cite{Antonio:2007pb}, the RBC collaboration
estimate $B_K$ with two and three dynamical flavors using the domain-wall
lattice fermion formalism.
While no clear dependence on the number of dynamical flavors is seen
between these two calculations, we cannot draw a definite conclusion
at the moment as these two calculations used different gauge actions.
Therefore, we leave the estimate of the systematic error due to the
missing strange quark contribution to the sea for future works.

Summing up the estimates of the systematic errors due to finite volume
effects (5 \%), fixing topology (1.4 \%), the matching factor (2 \%), and
the scale setting (5 \%) in quadrature, we quote our result of the
$N_f$=2 calculation obtained at $1/a\sim$ 1.67 GeV as
\begin{eqnarray}
   B_K^{\overline{\rm MS}}(2~\mbox{GeV})
 = 0.537(4)(40),
\label{eq:bk-final}
\end{eqnarray}
where the first and the second errors are statistical and systematic.
Notice that the systematic error shown here does not include those due
to the scaling violation and neglecting the dynamical strange quark.

\section{Summary}
\label{summary}

We performed a dynamical overlap fermion calculation of $B_K$ for the
first time.
Although the three-point functions are contaminated by the
wrapping-around kaons and the excited states because of the short
temporal extent of our lattice, thanks to the low-mode averaging the
statistical signal is substantially improved so that we could extract
the meson and antimeson transition amplitude accurately.

Using the extracted values of the $B$-parameter, consistency with the
NLO PQChPT prediction for $B_K$ is tested.
It turns out that the NLO prediction well describes the measured $B_P$
up to around a half of the strange quark mass.
By extrapolating $B_P$ to the physical $m_K$, we
obtain~(\ref{eq:bk-final}), where the uncertainties from the ordinary
finite volume effect, fixing topology and renormalization constant are
included in the systematic error.

The next step to do is the determination of $B_K$ in three-flavor QCD.
The generation of configurations with three flavors of dynamical overlap
fermions is underway~\cite{Hashimoto:2007vv}.
With this calculation, the effect of quenching the strange quark is
removed.
We are planning a study of the finite volume effects in the three-flavor
calculation by performing the calculation with two different volumes.
Then the dominant uncertainties in this calculation would be eliminated.

\section{Acknowledgment}

We would like to thank Damir Becirevic for giving us the numerical data
of~\cite{Becirevic:2003wk} and Dr. Enno Scholz for a useful comment.
H.F. would like to thank Nishina Memorial Foundation.
Numerical simulations are performed on the IBM System Blue Gene Solution
at High Energy Accelerator Research Organization (KEK) under support
of its Large Scale Simulation Program (No. 07-16).
This work is supported in part by the Grant-in-Aid of the Ministry of
Education (Nos. 17740171, 18034011, 18340075, 18740167, 18840045,
19540286, and 19740160).

\newpage

\clearpage
\begin{table}
 \begin{tabular}{cc|cccccc}
  $m_{v1}$ &
  $m_{v2}$ &
  $m_P$ &
  $Z_{A_4}^{\rm wall}f_P/2$ &
  $c_0$ &
  $c_1$ &
  $\Delta_P$ &
  $B_P^{\overline{\rm MS}}(2 {\rm GeV})$\\
  \hline
 0.015 & 0.015 & 0.172(1)  & 0.283(4) & 0.106(4) & 0.23(2) & 0.02(1)  & 0.43(1)\\
 0.025 & 0.015 & 0.197(1)  & 0.341(4) & 0.164(5) & 0.34(2) & 0.02(1)  & 0.458(7)\\
 0.025 & 0.025 & 0.220(1)  & 0.399(4) & 0.235(6) & 0.49(3) & 0.019(9) & 0.480(6)\\
 0.035 & 0.015 & 0.220(1)  & 0.394(4) & 0.228(6) & 0.48(3) & 0.02(1)  & 0.478(6)\\
 0.035 & 0.025 & 0.240(1)  & 0.451(4) & 0.311(7) & 0.65(3) & 0.021(8) & 0.497(5)\\
 0.035 & 0.035 & 0.259(1)  & 0.504(4) & 0.400(8) & 0.84(4) & 0.022(8) & 0.512(4)\\
 0.050 & 0.015 & 0.250(1)  & 0.466(5) & 0.334(8) & 0.69(4) & 0.021(9) & 0.501(5)\\
 0.050 & 0.025 & 0.268(1)  & 0.523(5) & 0.435(8) & 0.91(5) & 0.023(8) & 0.518(4)\\
 0.050 & 0.035 & 0.285(1)  & 0.575(5) & 0.540(9) & 1.14(5) & 0.024(7) & 0.530(3)\\
 0.050 & 0.050 & 0.3088(9) & 0.649(5) & 0.71(1)  & 1.50(6) & 0.025(7) & 0.545(3)\\
 0.070 & 0.015 & 0.285(1)  & 0.553(5) & 0.49(1)  & 1.02(6) & 0.02(1)  & 0.525(4)\\
 0.070 & 0.025 & 0.301(1)  & 0.610(5) & 0.62(1)  & 1.29(6) & 0.025(8) & 0.539(3)\\
 0.070 & 0.035 & 0.317(1)  & 0.663(5) & 0.74(1)  & 1.56(7) & 0.025(7) & 0.549(3)\\
 0.070 & 0.050 & 0.3384(9) & 0.737(5) & 0.94(1)  & 1.98(8) & 0.025(7) & 0.561(2)\\
 0.070 & 0.070 & 0.3659(9) & 0.829(5) & 1.22(2)  & 2.6(1)  & 0.025(7) & 0.575(2)\\
 0.100 & 0.015 & 0.332(1)  & 0.668(7) & 0.76(2)  & 1.6(1)  & 0.02(1)  & 0.552(4)\\
 0.100 & 0.025 & 0.346(1)  & 0.727(6) & 0.92(2)  & 1.9(1)  & 0.026(9) & 0.564(3)\\
 0.100 & 0.035 & 0.360(1)  & 0.780(6) & 1.07(2)  & 2.2(1)  & 0.026(8) & 0.572(3)\\
 0.100 & 0.050 & 0.3792(9) & 0.855(6) & 1.31(2)  & 2.7(1)  & 0.025(7) & 0.581(2)\\
 0.100 & 0.070 & 0.4044(8) & 0.948(6) & 1.64(2)  & 3.4(1)  & 0.023(7) & 0.593(2)\\
 0.100 & 0.100 & 0.4403(8) & 1.077(6) & 2.17(3)  & 4.4(2)  & 0.020(8) & 0.608(2)\\
  \hline
 \end{tabular}
 \caption{Numerical results at $m_{\rm sea}$=0.015 in the $Q=0$ sector.}
 \label{tab:mass-results-0.015}
\end{table}
\begin{table}
 \begin{tabular}{cc|cccccc}
  $m_{v1}$ &
  $m_{v2}$ &
  $m_P$ &
  $Z_{A_4}^{\rm wall}f_P/2$ &
  $c_0$ &
  $c_1$ &
  $\Delta_P$ &
  $B_P^{\overline{\rm MS}}(2 {\rm GeV})$\\
  \hline
 0.015 & 0.015 & 0.170(1)  & 0.314(4) & 0.129(4) & 0.28(2) & 0.016(9) & 0.43(1)\\
 0.025 & 0.015 & 0.1963(9) & 0.373(4) & 0.194(5) & 0.43(2) & 0.023(8) & 0.454(7)\\
 0.025 & 0.025 & 0.2190(9) & 0.430(4) & 0.272(5) & 0.60(3) & 0.027(7) & 0.477(5)\\
 0.035 & 0.015 & 0.2189(9) & 0.427(4) & 0.265(6) & 0.58(3) & 0.028(8) & 0.474(6)\\
 0.035 & 0.025 & 0.2395(9) & 0.482(4) & 0.354(6) & 0.78(4) & 0.031(7) & 0.494(4)\\
 0.035 & 0.035 & 0.2585(8) & 0.535(4) & 0.447(7) & 1.00(4) & 0.033(7) & 0.509(4)\\
 0.050 & 0.015 & 0.2492(9) & 0.500(4) & 0.382(7) & 0.85(4) & 0.033(8) & 0.496(5)\\
 0.050 & 0.025 & 0.2675(9) & 0.555(4) & 0.486(7) & 1.09(5) & 0.034(7) & 0.514(4)\\
 0.050 & 0.035 & 0.2846(8) & 0.606(4) & 0.594(8) & 1.34(6) & 0.036(7) & 0.527(3)\\
 0.050 & 0.050 & 0.3087(8) & 0.678(4) & 0.766(9) & 1.74(8) & 0.037(7) & 0.542(3)\\
 0.070 & 0.015 & 0.2848(9) & 0.589(5) & 0.55(1)  & 1.25(7) & 0.038(9) & 0.519(4)\\
 0.070 & 0.025 & 0.3010(9) & 0.643(4) & 0.68(1)  & 1.53(7) & 0.037(7) & 0.534(3)\\
 0.070 & 0.035 & 0.3165(8) & 0.693(4) & 0.80(1)  & 1.83(8) & 0.037(7) & 0.545(3)\\
 0.070 & 0.050 & 0.3384(8) & 0.764(4) & 1.00(1)  & 2.3(1)  & 0.038(7) & 0.559(2)\\
 0.070 & 0.070 & 0.3661(8) & 0.854(4) & 1.28(1)  & 2.9(1)  & 0.038(7) & 0.573(2)\\
 0.100 & 0.015 & 0.332(1)  & 0.708(6) & 0.84(1)  & 1.9(1)  & 0.04(1)  & 0.545(4)\\
 0.100 & 0.025 & 0.3461(9) & 0.761(5) & 0.99(1)  & 2.2(1)  & 0.038(8) & 0.559(3)\\
 0.100 & 0.035 & 0.3597(9) & 0.811(5) & 1.15(1)  & 2.6(1)  & 0.038(7) & 0.568(2)\\
 0.100 & 0.050 & 0.3795(8) & 0.881(5) & 1.38(2)  & 3.1(1)  & 0.038(7) & 0.579(2)\\
 0.100 & 0.070 & 0.4048(8) & 0.971(5) & 1.71(2)  & 3.9(2)  & 0.038(7) & 0.591(2)\\
 0.100 & 0.100 & 0.4408(8) & 1.096(6) & 2.24(3)  & 5.1(2)  & 0.038(7) & 0.606(2)\\
  \hline
 \end{tabular}
 \caption{Numerical results at $m_{\rm sea}$=0.025 in the $Q=0$ sector.}
 \label{tab:mass-results-0.025}
\end{table}
\begin{table}
 \begin{tabular}{cc|cccccc}
  $m_{v1}$ &
  $m_{v2}$ &
  $m_P$ &
  $Z_{A_4}^{\rm wall}f_P/2$ &
  $c_0$ &
  $c_1$ &
  $\Delta_P$ &
  $B_P^{\overline{\rm MS}}(2 {\rm GeV})$\\
  \hline
 0.015 & 0.015 & 0.173(1)  & 0.325(3) & 0.138(5) & 0.31(2) & 0.02(1)  & 0.43(1)\\
 0.025 & 0.015 & 0.199(1)  & 0.382(4) & 0.203(5) & 0.44(2) & 0.019(9) & 0.451(7)\\
 0.025 & 0.025 & 0.2211(9) & 0.437(4) & 0.279(6) & 0.60(3) & 0.021(7) & 0.474(6)\\
 0.035 & 0.015 & 0.221(1)  & 0.436(4) & 0.274(6) & 0.59(3) & 0.021(8) & 0.470(6)\\
 0.035 & 0.025 & 0.2416(9) & 0.488(4) & 0.360(7) & 0.77(3) & 0.023(7) & 0.491(5)\\
 0.035 & 0.035 & 0.2604(9) & 0.538(4) & 0.452(7) & 0.98(4) & 0.025(6) & 0.507(4)\\
 0.050 & 0.015 & 0.251(1)  & 0.509(4) & 0.392(7) & 0.84(4) & 0.024(8) & 0.492(5)\\
 0.050 & 0.025 & 0.2695(9) & 0.559(4) & 0.491(8) & 1.06(4) & 0.025(6) & 0.511(4)\\
 0.050 & 0.035 & 0.2864(9) & 0.608(4) & 0.596(9) & 1.29(5) & 0.027(6) & 0.525(4)\\
 0.050 & 0.050 & 0.3104(8) & 0.679(4) & 0.77(1)  & 1.66(6) & 0.028(6) & 0.540(3)\\
 0.070 & 0.015 & 0.287 (1) & 0.599(5) & 0.57(1)  & 1.20(6) & 0.025(9) & 0.514(4)\\
 0.070 & 0.025 & 0.3030(9) & 0.647(4) & 0.68(1)  & 1.47(6) & 0.027(7) & 0.531(4)\\
 0.070 & 0.035 & 0.3182(9) & 0.694(4) & 0.80(1)  & 1.74(7) & 0.028(6) & 0.543(3)\\
 0.070 & 0.050 & 0.3401(8) & 0.763(4) & 1.00(1)  & 2.15(8) & 0.028(6) & 0.556(3)\\
 0.070 & 0.070 & 0.3676(8) & 0.852(5) & 1.27(1)  & 2.7(1)  & 0.026(6) & 0.571(3)\\
 0.100 & 0.015 & 0.334(1)  & 0.719(6) & 0.86(1)  & 1.8(1)  & 0.02(1)  & 0.538(4)\\
 0.100 & 0.025 & 0.3479(9) & 0.765(5) & 1.00(1)  & 2.1(1)  & 0.026(7) & 0.554(3)\\
 0.100 & 0.035 & 0.3614(9) & 0.811(5) & 1.14(1)  & 2.4(1)  & 0.027(6) & 0.564(3)\\
 0.100 & 0.050 & 0.3810(8) & 0.879(5) & 1.37(2)  & 2.9(1)  & 0.026(6) & 0.576(3)\\
 0.100 & 0.070 & 0.4062(8) & 0.967(5) & 1.69(2)  & 3.5(1)  & 0.023(6) & 0.589(3)\\
 0.100 & 0.100 & 0.4422(8) & 1.091(6) & 2.21(2)  & 4.5(2)  & 0.018(7) & 0.605(2)\\
  \hline
 \end{tabular}
 \caption{Numerical results at $m_{\rm sea}$=0.035 in the $Q=0$ sector.}
 \label{tab:mass-results-0.035}
\end{table}
\begin{table}
 \begin{tabular}{cc|cccccc}
  $m_{v1}$ &
  $m_{v2}$ &
  $m_P$ &
  $Z_{A_4}^{\rm wall}f_P/2$ &
  $c_0$ &
  $c_1$ &
  $\Delta_P$ &
  $B_P^{\overline{\rm MS}}(2 {\rm GeV})$\\
  \hline
 0.015 & 0.015 & 0.175(1)  & 0.350(3) & 0.162(4) & 0.34(2) & 0.005(9) & 0.430(8)\\
 0.025 & 0.015 & 0.2006(9) & 0.409(3) & 0.236(5) & 0.49(2) & 0.011(8) & 0.458(6)\\
 0.025 & 0.025 & 0.2232(9) & 0.464(3) & 0.319(5) & 0.67(3) & 0.016(7) & 0.481(5)\\
 0.035 & 0.015 & 0.2234(9) & 0.464(3) & 0.317(6) & 0.66(3) & 0.016(8) & 0.478(5)\\
 0.035 & 0.025 & 0.2437(9) & 0.515(3) & 0.406(6) & 0.86(4) & 0.020(6) & 0.498(4)\\
 0.035 & 0.035 & 0.2626(9) & 0.566(3) & 0.504(7) & 1.07(4) & 0.023(6) & 0.512(4)\\
 0.050 & 0.015 & 0.2538(9) & 0.540(4) & 0.449(7) & 0.95(5) & 0.022(8) & 0.501(4)\\
 0.050 & 0.025 & 0.2718(9) & 0.588(4) & 0.549(8) & 1.17(5) & 0.024(6) & 0.517(4)\\
 0.050 & 0.035 & 0.2888(9) & 0.635(4) & 0.657(9) & 1.41(6) & 0.026(6) & 0.529(3)\\
 0.050 & 0.050 & 0.3128(9) & 0.705(4) & 0.83(1)  & 1.79(7) & 0.028(6) & 0.544(3)\\
 0.070 & 0.015 & 0.290(1)  & 0.632(4) & 0.64(1)  & 1.38(8) & 0.027(9) & 0.524(4)\\
 0.070 & 0.025 & 0.3054(9) & 0.677(4) & 0.76(1)  & 1.63(7) & 0.028(7) & 0.537(3)\\
 0.070 & 0.035 & 0.3207(9) & 0.722(4) & 0.88(1)  & 1.90(8) & 0.029(6) & 0.547(3)\\
 0.070 & 0.050 & 0.3426(9) & 0.789(4) & 1.07(1)  & 2.33(9) & 0.029(6) & 0.560(3)\\
 0.070 & 0.070 & 0.3703(8) & 0.876(4) & 1.36(2)  & 2.9(1)  & 0.030(6) & 0.575(3)\\
 0.100 & 0.015 & 0.337(1)  & 0.755(5) & 0.96(2)  & 2.1(1)  & 0.03(1)  & 0.549(4)\\
 0.100 & 0.025 & 0.351(1)  & 0.798(5) & 1.10(1)  & 2.4(1)  & 0.032(8) & 0.560(3)\\
 0.100 & 0.035 & 0.3641(9) & 0.841(5) & 1.24(2)  & 2.7(1)  & 0.031(7) & 0.569(3)\\
 0.100 & 0.050 & 0.3838(9) & 0.906(5) & 1.47(2)  & 3.2(1)  & 0.031(7) & 0.580(3)\\
 0.100 & 0.070 & 0.4091(9) & 0.991(5) & 1.79(2)  & 3.9(2)  & 0.031(7) & 0.593(3)\\
 0.100 & 0.100 & 0.4452(9) & 1.114(6) & 2.32(3)  & 5.1(2)  & 0.032(7) & 0.608(3)\\
  \hline
 \end{tabular}
 \caption{Numerical results at $m_{\rm sea}$=0.050 in the $Q=0$ sector.}
 \label{tab:mass-results-0.050}
\end{table}
\begin{table}
 \begin{tabular}{cc|cccccc}
  $m_{v1}$ &
  $m_{v2}$ &
  $m_P$ &
  $Z_{A_4}^{\rm wall}f_P/2$ &
  $c_0$ &
  $c_1$ &
  $\Delta_P$ &
  $B_P^{\overline{\rm MS}}(2 {\rm GeV})$\\
  \hline
 0.015 & 0.015 & 0.1760(9) & 0.365(4) & 0.178(5) & 0.34(2) & $-$0.005(9) & 0.436(8)\\
 0.025 & 0.015 & 0.2018(8) & 0.423(4) & 0.254(6) & 0.49(3) &    0.005(8) & 0.462(6)\\
 0.025 & 0.025 & 0.2243(8) & 0.476(4) & 0.338(6) & 0.68(3) &    0.013(7) & 0.484(5)\\
 0.035 & 0.015 & 0.2245(8) & 0.477(4) & 0.337(7) & 0.67(3) &    0.012(8) & 0.481(5)\\
 0.035 & 0.025 & 0.2448(8) & 0.526(4) & 0.426(7) & 0.88(4) &    0.020(7) & 0.500(4)\\
 0.035 & 0.035 & 0.2636(8) & 0.575(4) & 0.522(8) & 1.11(5) &    0.026(6) & 0.514(4)\\
 0.050 & 0.015 & 0.2547(8) & 0.552(4) & 0.472(9) & 0.98(5) &    0.021(8) & 0.503(4)\\
 0.050 & 0.025 & 0.2728(8) & 0.597(4) & 0.569(9) & 1.22(5) &    0.027(7) & 0.519(4)\\
 0.050 & 0.035 & 0.2898(7) & 0.642(4) & 0.67(1)  & 1.48(6) &    0.032(6) & 0.531(3)\\
 0.050 & 0.050 & 0.3139(7) & 0.710(4) & 0.85(1)  & 1.90(8) &    0.036(6) & 0.546(3)\\
 0.070 & 0.015 & 0.2903(8) & 0.643(5) & 0.67(1)  & 1.46(9) &    0.031(9) & 0.525(4)\\
 0.070 & 0.025 & 0.3064(8) & 0.684(4) & 0.78(1)  & 1.72(8) &    0.034(7) & 0.539(3)\\
 0.070 & 0.035 & 0.3217(7) & 0.727(4) & 0.89(1)  & 2.02(8) &    0.037(6) & 0.549(3)\\
 0.070 & 0.050 & 0.3437(7) & 0.793(4) & 1.09(1)  & 2.5(1)  &    0.039(6) & 0.562(2)\\
 0.070 & 0.070 & 0.3714(7) & 0.878(5) & 1.36(2)  & 3.1(1)  &    0.039(6) & 0.576(2)\\
 0.100 & 0.015 & 0.3373(9) & 0.766(6) & 0.99(2)  & 2.3(2)  &    0.04(1)  & 0.551(4)\\
 0.100 & 0.025 & 0.3514(8) & 0.804(5) & 1.12(2)  & 2.6(1)  &    0.043(8) & 0.563(3)\\
 0.100 & 0.035 & 0.3651(8) & 0.845(5) & 1.25(2)  & 2.9(1)  &    0.042(7) & 0.571(3)\\
 0.100 & 0.050 & 0.3849(7) & 0.908(5) & 1.47(2)  & 3.4(1)  &    0.041(6) & 0.582(2)\\
 0.100 & 0.070 & 0.4103(7) & 0.991(5) & 1.79(2)  & 4.1(2)  &    0.040(6) & 0.594(2)\\
 0.100 & 0.100 & 0.4464(7) & 1.112(6) & 2.32(2)  & 5.3(2)  &    0.039(6) & 0.609(2)\\
  \hline
 \end{tabular}
 \caption{Numerical results at $m_{\rm sea}$=0.070 in the $Q=0$ sector.}
 \label{tab:mass-results-0.070}
\end{table}
\begin{table}
 \begin{tabular}{cc|cccccc}
  $m_{v1}$ &
  $m_{v2}$ &
  $m_P$ &
  $Z_{A_4}^{\rm wall}f_P/2$ &
  $c_0$ &
  $c_1$ &
  $\Delta_P$ &
  $B_P^{\overline{\rm MS}}(2 {\rm GeV})$\\
  \hline
 0.015 & 0.015 & 0.1770(8) & 0.374(3) & 0.195(5) & 0.42(3) & 0.02(1)  & 0.454(8)\\
 0.025 & 0.015 & 0.2032(8) & 0.434(4) & 0.276(5) & 0.58(3) & 0.018(8) & 0.476(6)\\
 0.025 & 0.025 & 0.2260(7) & 0.490(3) & 0.364(6) & 0.75(3) & 0.018(7) & 0.494(4)\\
 0.035 & 0.015 & 0.2261(8) & 0.489(4) & 0.362(6) & 0.75(4) & 0.020(7) & 0.491(5)\\
 0.035 & 0.025 & 0.2467(7) & 0.541(4) & 0.456(7) & 0.94(4) & 0.020(6) & 0.507(4)\\
 0.035 & 0.035 & 0.2657(7) & 0.591(4) & 0.557(8) & 1.15(5) & 0.021(6) & 0.519(3)\\
 0.050 & 0.015 & 0.2566(8) & 0.565(4) & 0.500(8) & 1.05(5) & 0.023(7) & 0.509(4)\\
 0.050 & 0.025 & 0.2749(7) & 0.613(4) & 0.604(8) & 1.25(5) & 0.022(6) & 0.523(3)\\
 0.050 & 0.035 & 0.2920(7) & 0.659(4) & 0.713(9) & 1.48(6) & 0.023(6) & 0.533(3)\\
 0.050 & 0.050 & 0.3161(7) & 0.727(4) & 0.89(1)  & 1.86(7) & 0.024(6) & 0.546(2)\\
 0.070 & 0.015 & 0.2925(8) & 0.657(5) & 0.70(1)  & 1.49(7) & 0.027(7) & 0.528(4)\\
 0.070 & 0.025 & 0.3086(7) & 0.701(4) & 0.82(1)  & 1.71(7) & 0.026(6) & 0.540(3)\\
 0.070 & 0.035 & 0.3240(7) & 0.744(4) & 0.94(1)  & 1.97(8) & 0.025(6) & 0.550(2)\\
 0.070 & 0.050 & 0.3459(7) & 0.809(4) & 1.13(1)  & 2.38(9) & 0.026(6) & 0.561(2)\\
 0.070 & 0.070 & 0.3735(7) & 0.894(4) & 1.41(1)  & 3.0(1)  & 0.026(6) & 0.574(2)\\
 0.100 & 0.015 & 0.3397(9) & 0.780(6) & 1.03(2)  & 2.2(1)  & 0.032(9) & 0.550(4)\\
 0.100 & 0.025 & 0.3538(8) & 0.821(5) & 1.16(2)  & 2.5(1)  & 0.029(7) & 0.561(3)\\
 0.100 & 0.035 & 0.3674(7) & 0.862(5) & 1.30(1)  & 2.8(1)  & 0.028(6) & 0.569(2)\\
 0.100 & 0.050 & 0.3871(7) & 0.924(5) & 1.52(2)  & 3.2(1)  & 0.027(6) & 0.579(2)\\
 0.100 & 0.070 & 0.4123(7) & 1.005(5) & 1.83(2)  & 3.9(2)  & 0.026(6) & 0.590(2)\\
 0.100 & 0.100 & 0.4482(7) & 1.123(6) & 2.34(2)  & 4.9(2)  & 0.026(7) & 0.604(1)\\
  \hline
 \end{tabular}
 \caption{Numerical results at $m_{\rm sea}$=0.100 in the $Q=0$ sector.}
 \label{tab:mass-results-0.100}
\end{table}
\begin{table}
 \begin{tabular}{cc|cccccc}
  $m_{v1}$ &
  $m_{v2}$ &
  $m_P$ &
  $Z_{A_4}^{\rm wall}f_P/2$ &
  $c_0$ &
  $c_1$ &
  $\Delta_P$ &
  $B_P^{\overline{\rm MS}}(2 {\rm GeV})$\\
  \hline
 0.015 & 0.015 & 0.177(1) & 0.351(5) & 0.169(6) & 0.33(3) & 0.00(1)  & 0.446(9)\\
 0.025 & 0.015 & 0.201(1) & 0.410(5) & 0.240(8) & 0.46(4) & 0.00(1)  & 0.464(7)\\
 0.025 & 0.025 & 0.223(1) & 0.464(5) & 0.317(9) & 0.61(4) & 0.00(1)  & 0.479(7)\\
 0.035 & 0.015 & 0.224(1) & 0.467(6) & 0.32(1)  & 0.62(4) & 0.01(1)  & 0.480(7)\\
 0.035 & 0.025 & 0.244(1) & 0.516(5) & 0.40(1)  & 0.78(5) & 0.01(1)  & 0.492(7)\\
 0.035 & 0.035 & 0.262(1) & 0.567(5) & 0.50(1)  & 0.96(6) & 0.007(9) & 0.504(6)\\
 0.050 & 0.015 & 0.254(1) & 0.545(6) & 0.46(1)  & 0.88(6) & 0.01(1)  & 0.500(7)\\
 0.050 & 0.025 & 0.271(1) & 0.590(6) & 0.55(1)  & 1.05(6) & 0.01(1)  & 0.509(6)\\
 0.050 & 0.035 & 0.288(1) & 0.638(6) & 0.65(1)  & 1.25(7) & 0.008(9) & 0.520(6)\\
 0.050 & 0.050 & 0.312(1) & 0.710(6) & 0.83(2)  & 1.57(9) & 0.007(9) & 0.535(6)\\
 0.070 & 0.015 & 0.289(1) & 0.641(7) & 0.66(2)  & 1.28(9) & 0.01(1)  & 0.523(7)\\
 0.070 & 0.025 & 0.305(1) & 0.681(6) & 0.76(2)  & 1.45(9) & 0.01(1)  & 0.529(6)\\
 0.070 & 0.035 & 0.320(1) & 0.727(6) & 0.88(2)  & 1.7(1)  & 0.01(1)  & 0.539(6)\\
 0.070 & 0.050 & 0.342(1) & 0.796(6) & 1.08(2)  & 2.0(1)  & 0.01(1)  & 0.552(6)\\
 0.070 & 0.070 & 0.370(1) & 0.886(7) & 1.37(3)  & 2.5(2)  & 0.00(1)  & 0.568(6)\\
 0.100 & 0.015 & 0.336(1) & 0.769(8) & 1.00(3)  & 2.0(2)  & 0.02(2)  & 0.550(8)\\
 0.100 & 0.025 & 0.350(1) & 0.805(7) & 1.10(2)  & 2.1(1)  & 0.01(1)  & 0.554(7)\\
 0.100 & 0.035 & 0.363(1) & 0.849(7) & 1.25(3)  & 2.3(2)  & 0.01(1)  & 0.562(7)\\
 0.100 & 0.050 & 0.383(1) & 0.916(7) & 1.48(3)  & 2.7(2)  & 0.00(1)  & 0.574(6)\\
 0.100 & 0.070 & 0.409(1) & 1.003(7) & 1.82(3)  & 3.3(2)  & 0.00(1)  & 0.588(6)\\
 0.100 & 0.100 & 0.444(1) & 1.127(8) & 2.36(4)  & 4.2(3)  & 0.00(1)  & 0.605(5)\\
  \hline
 \end{tabular}
 \caption{Numerical results at $m_{\rm sea}$=0.050 in the $Q=-2$ sector.}
 \label{tab:mass-results-0.050-Q-2}
\end{table}
\begin{table}
 \begin{tabular}{cc|cccccc}
  $m_{v1}$ &
  $m_{v2}$ &
  $m_P$ &
  $Z_{A_4}^{\rm wall}f_P/2$ &
  $c_0$ &
  $c_1$ &
  $\Delta_P$ &
  $B_P^{\overline{\rm MS}}(2 {\rm GeV})$\\
  \hline
 0.015 & 0.015 & 0.185(1)  & 0.378(4) & 0.198(6) & 0.44(3) & 0.02(1)  & 0.451(8)\\
 0.025 & 0.015 & 0.208(1)  & 0.436(5) & 0.278(7) & 0.62(4) & 0.022(9) & 0.476(6)\\
 0.025 & 0.025 & 0.229(1)  & 0.482(5) & 0.349(8) & 0.77(4) & 0.024(8) & 0.489(5)\\
 0.035 & 0.015 & 0.230(1)  & 0.494(5) & 0.373(9) & 0.82(5) & 0.024(9) & 0.497(5)\\
 0.035 & 0.025 & 0.2480(9) & 0.531(5) & 0.437(9) & 0.96(5) & 0.026(8) & 0.504(4)\\
 0.035 & 0.035 & 0.2658(9) & 0.578(5) & 0.53(1)  & 1.17(6) & 0.028(8) & 0.515(4)\\
 0.050 & 0.015 & 0.259(1)  & 0.576(5) & 0.53(1)  & 1.17(7) & 0.027(8) & 0.522(5)\\
 0.050 & 0.025 & 0.275(1)  & 0.604(5) & 0.59(1)  & 1.30(7) & 0.029(8) & 0.523(4)\\
 0.050 & 0.035 & 0.291(1)  & 0.645(5) & 0.68(1)  & 1.51(8) & 0.029(8) & 0.532(4)\\
 0.050 & 0.050 & 0.314(1)  & 0.711(5) & 0.85(1)  & 1.9(1)  & 0.030(9) & 0.546(4)\\
 0.070 & 0.015 & 0.294(1)  & 0.676(6) & 0.77(2)  & 1.7(1)  & 0.032(9) & 0.547(5)\\
 0.070 & 0.025 & 0.308(1)  & 0.695(6) & 0.81(2)  & 1.8(1)  & 0.032(9) & 0.545(4)\\
 0.070 & 0.035 & 0.322(1)  & 0.732(6) & 0.91(2)  & 2.0(1)  & 0.031(9) & 0.551(4)\\
 0.070 & 0.050 & 0.343(1)  & 0.794(6) & 1.09(2)  & 2.4(1)  & 0.03(1)  & 0.563(5)\\
 0.070 & 0.070 & 0.371(1)  & 0.878(6) & 1.37(2)  & 3.0(2)  & 0.03(1)  & 0.577(5)\\
 0.100 & 0.015 & 0.340(1)  & 0.810(7) & 1.16(2)  & 2.7(2)  & 0.04(1)  & 0.575(5)\\
 0.100 & 0.025 & 0.352(1)  & 0.820(6) & 1.18(2)  & 2.7(2)  & 0.04(1)  & 0.569(5)\\
 0.100 & 0.035 & 0.365(1)  & 0.852(6) & 1.28(2)  & 2.9(2)  & 0.03(1)  & 0.574(5)\\
 0.100 & 0.050 & 0.384(1)  & 0.910(7) & 1.48(2)  & 3.2(2)  & 0.03(1)  & 0.583(5)\\
 0.100 & 0.070 & 0.409(1)  & 0.990(7) & 1.79(3)  & 3.8(3)  & 0.02(1)  & 0.595(5)\\
 0.100 & 0.100 & 0.444(1)  & 1.110(8) & 2.31(4)  & 4.7(4)  & 0.02(1)  & 0.610(5)\\
  \hline
 \end{tabular}
 \caption{Numerical results at $m_{\rm sea}$=0.050 in the $Q=-4$ sector.}
 \label{tab:mass-results-0.050-Q-4}
\end{table}
\newcommand{\cc}[1]{\multicolumn{6}{c}{#1}}
\begin{table}
 \begin{tabular}{c|ccccc}
  Fit range in $m_{\rm sea}$ & $\chi^2$/dof & $B_P^\chi$ & $b_1-b_3$
                             & $b_2$ & $f$\\
  \hline
  \cc{Fit results with unfixed $f$}\\
  \hline
  $[0.015,\ 0.035]$ & 0.21 & 0.26(6) & 0.3(6)  & $-$0.2(3)
                    & 0.06(2)\\
  $[0.015,\ 0.050]$ & 0.25 & 0.31(3) & 0.9(1)  & $-$0.1(1)
                    & 0.09(2)\\
  $[0.015,\ 0.070]$ & 0.33 & 0.33(2) & 1.03(3) & $-$0.09(8)
                    & 0.101(8)\\
  $[0.015,\ 0.100]$ & 0.75 & 0.36(1) & 1.04(3) & $-$0.05(4)
                    & 0.120(6)\\
  \hline
  \cc{Fit results with $f$ fixed to 0.0659}\\
  \hline
  $[0.015,\ 0.035]$ & 0.15 & 0.27(1) & 0.5(2) & $-$0.2(3) & -\\
  $[0.015,\ 0.050]$ & 0.46 & 0.260(5)& 0.76(6) & $-$0.1(1) & -\\
  $[0.015,\ 0.070]$ & 2.5  & 0.246(3)& 1.07(3) & $-$0.09(8)& -\\
  $[0.015,\ 0.100]$ &13.5  & 0.224(2)& 1.42(2) & $-$0.03(5)& -\\
 \end{tabular}
 \caption{Results of the NLO PQChPT fit.
 In both fits, $\mu$ is set to 1 GeV.}
 \label{tab:nlo-test}
\end{table}
\renewcommand{\cc}[1]{\multicolumn{9}{c}{#1}}
\begin{table}
 \begin{tabular}{c|cccccccc}
  Fit range in $m_{\rm sea}$ & $\chi^2$/dof & $B_P^\chi$ & $b_1$ & $b_2$
& $b_3$ & $d_1$ & $f$ & $B_K^{\overline{\rm MS}}(2\,{\rm GeV})$\\
  \hline
  \cc{Fit results with unfixed $f$}\\
  \hline
  $[0.015,\ 0.035]$ & 0.1        & 0.31(2) & $-$0.5(3) & $-$0.15(10)
                    & $-$0.11(3) &  4.0(9) &  0.079(8) & 0.539(7)\\
  $[0.015,\ 0.050]$ & 0.4        & 0.31(2) & $-$0.4(3) & $-$0.06(5)
                    & $-$0.10(3) & 3.9(1.0)&  0.080(9) & 0.537(4)\\
  $[0.015,\ 0.070]$ & 0.5        & 0.31(2) & $-$0.4(3) & $-$0.02(3)
                    & $-$0.09(4) & 3.9(1.0)&  0.081(9) & 0.536(4)\\
  $[0.015,\ 0.100]$ & 0.5        & 0.31(2) & $-$0.3(3) & $-$0.03(2)
                    & $-$0.09(4) & 3.5(1.2)& 0.083(10) & 0.535(4)\\
 \end{tabular}
 \caption{Results of fitting to the NLO PQChPT plus a quadratic term.
 In the fits, $\mu$ is set to 1 GeV.}
 \label{tab:nnlo}
\end{table}


\begin{thebibliography}{99}

 \bibitem{Yao:2006px}
  W.~M.~Yao {\it et al.}  [Particle Data Group],
  J.\ Phys.\ G {\bf 33}, 1 (2006).

\bibitem{Buchalla:1995vs}
  For the full expression, see
  G.~Buchalla, A.~J.~Buras and M.~E.~Lautenbacher,
  Rev.\ Mod.\ Phys.\  {\bf 68}, 1125 (1996)
  [arXiv:hep-ph/9512380].

\bibitem{Flynn:2004au}
  J.~M.~Flynn, F.~Mescia and A.~S.~B.~Tariq  [UKQCD Collaboration],
  JHEP {\bf 0411}, 049 (2004)
  [arXiv:hep-lat/0406013].

\bibitem{Mescia:2005ew}
  F.~Mescia, V.~Gimenez, V.~Lubicz, G.~Martinelli, S.~Simula and C.~Tarantino,
  PoS {\bf LAT2005}, 365 (2006)
  [arXiv:hep-lat/0510096].

\bibitem{Aoki:1997nr}
  S.~Aoki {\it et al.}  [JLQCD Collaboration],
  Phys.\ Rev.\ Lett.\  {\bf 80}, 5271 (1998)
  [arXiv:hep-lat/9710073].

 \bibitem{AliKhan:2001wr}
  A.~Ali Khan {\it et al.}  [CP-PACS Collaboration],
  Phys.\ Rev.\  D {\bf 64}, 114506 (2001)
  [arXiv:hep-lat/0105020].

 \bibitem{Blum:2001xb}
  T.~Blum {\it et al.}  [RBC Collaboration],
  Phys.\ Rev.\  D {\bf 68}, 114506 (2003)
  [arXiv:hep-lat/0110075].

\bibitem{Aoki:2005ga}
  Y.~Aoki {\it et al.},
  Phys.\ Rev.\  D {\bf 73}, 094507 (2006)
  [arXiv:hep-lat/0508011].

\bibitem{Antonio:2007pb}
  D.~J.~Antonio {\it et al.}  [RBC Collaboration],
  arXiv:hep-ph/0702042.

\bibitem{Aoki:2004ht}
  Y.~Aoki {\it et al.},
  Phys.\ Rev.\  D {\bf 72}, 114505 (2005)
  [arXiv:hep-lat/0411006].

\bibitem{Aoki:2007xm}
  Y.~Aoki {\it et al.},
  arXiv:0712.1061 [hep-lat].

\bibitem{Fodor:2003bh}
  Z.~Fodor, S.~D.~Katz and K.~K.~Szabo,
  JHEP {\bf 0408}, 003 (2004)
  [arXiv:hep-lat/0311010].

\bibitem{Fukaya:2006vs}
  H.~Fukaya, S.~Hashimoto, K.~I.~Ishikawa, T.~Kaneko, H.~Matsufuru,
  T.~Onogi and N.~Yamada [JLQCD Collaboration],
  Phys.\ Rev.\  D {\bf 74}, 094505 (2006)
  [arXiv:hep-lat/0607020].

\bibitem{Aoki:2007ka}
  S.~Aoki, H.~Fukaya, S.~Hashimoto and T.~Onogi,
  Phys.\ Rev.\  D {\bf 76}, 054508 (2007)
  [arXiv:0707.0396 [hep-lat]].

\bibitem{Noaki:2007es}
  J.~Noaki {\it et al.}  [JLQCD Collaboration],
  arXiv:0710.0929 [hep-lat].

\bibitem{Kaneko:2007nf}
  T.~Kaneko, H.~Fukaya, S.~Hashimoto, H.~Matsufuru, J.~Noaki, T.~Onogi
  and N.~Yamada [JLQCD collaboration],
  arXiv:0710.2390 [hep-lat].

\bibitem{Shintani:2007ub}
  E.~Shintani, H.~Fukaya, S.~Hashimoto, H.~Matsufuru, J.~Noaki, T.~Onogi
  and N.~Yamada [JLQCD Collaboration],
  arXiv:0710.0691 [hep-lat].

\bibitem{Aoki:2007pw}
  S.~Aoki {\it et al.}  [JLQCD and TWQCD Collaboration],
  arXiv:0710.1130 [hep-lat].

\bibitem{Fukaya:2007fb}
  H.~Fukaya {\it et al.}  [JLQCD Collaboration],
  Phys.\ Rev.\ Lett.\  {\bf 98}, 172001 (2007)
  [arXiv:hep-lat/0702003].

\bibitem{Fukaya:2007yv}
  H.~Fukaya {\it et al.},
  Phys.\ Rev.\  D {\bf 76}, 054503 (2007)
  [arXiv:0705.3322 [hep-lat]].

\bibitem{Fukaya:2007cw}
  H.~Fukaya {\it et al.}  [JLQCD collaboration],
  arXiv:0710.3468 [hep-lat].

\bibitem{Fukaya:2007pn}
  H.~Fukaya {\it et al.}  [JLQCD collaboration],
  arXiv:0711.4965 [hep-lat].

\bibitem{Matsufuru:2007uc}
  H.~Matsufuru  [JLQCD Collaboration],
  arXiv:0710.4225 [hep-lat].

\bibitem{Hashimoto:2007vv}
  S.~Hashimoto {\it et al.}  [JLQCD collaboration],
  arXiv:0710.2730 [hep-lat].

\bibitem{Neuberger:1997fp}
  H.~Neuberger,
  Phys.\ Lett.\  B {\bf 417}, 141 (1998)
  [arXiv:hep-lat/9707022].

\bibitem{Neuberger:1998wv}
  H.~Neuberger,
  Phys.\ Lett.\  B {\bf 427}, 353 (1998)
  [arXiv:hep-lat/9801031].

\bibitem{Sommer:1993ce}
  R.~Sommer,
  Nucl.\ Phys.\  B {\bf 411}, 839 (1994)
  [arXiv:hep-lat/9310022].

 \bibitem{Kaneko:2006pa}
  T.~Kaneko {\it et al.}  [JLQCD Collaboration],
  PoS {\bf LAT2006}, 054 (2006)
  [arXiv:hep-lat/0610036];
  S.~Aoki {\it et al.}  [JLQCD Collaboration], 
  arXiv:0803.3197.

\bibitem{DeGrand:2004qw}
  T.~A.~DeGrand and S.~Schaefer,
  Comput.\ Phys.\ Commun.\  {\bf 159}, 185 (2004)
  [arXiv:hep-lat/0401011];
  L.~Giusti, P.~Hernandez, M.~Laine, P.~Weisz and H.~Wittig,
  JHEP {\bf 0404}, 013 (2004)
  [arXiv:hep-lat/0402002].

\bibitem{Yagi:lat07}
  T.~Yagi, M.~Ohtani, O.~Morimatsu, S.~Hashimoto,
  talk given at the XXV International Symposium On Lattice Field Theory,
  University of Regensburg, July 30-August 4, 2007,
  PosLat (2007)086.

\bibitem{Martinelli:1994ty}
  G.~Martinelli, C.~Pittori, C.~T.~Sachrajda, M.~Testa and A.~Vladikas,
  Nucl.\ Phys.\  B {\bf 445}, 81 (1995)
  [arXiv:hep-lat/9411010];
  A.~Donini, V.~Gimenez, G.~Martinelli, M.~Talevi and A.~Vladikas,
  Eur.\ Phys.\ J.\  C {\bf 10}, 121 (1999)
  [arXiv:hep-lat/9902030].

\bibitem{Ciuchini:1997bw}
  M.~Ciuchini, E.~Franco, V.~Lubicz, G.~Martinelli, I.~Scimemi and L.~Silvestrini,
  Nucl.\ Phys.\  B {\bf 523}, 501 (1998)
  [arXiv:hep-ph/9711402].

 \bibitem{Della Morte:2004bc}
  M.~Della Morte, R.~Frezzotti, J.~Heitger, J.~Rolf, R.~Sommer and U.~Wolff
                  [ALPHA Collaboration],
  Nucl.\ Phys.\  B {\bf 713}, 378 (2005)
  [arXiv:hep-lat/0411025].

 \bibitem{Becirevic:2003wk}
  D.~Becirevic and G.~Villadoro,
  Phys.\ Rev.\  D {\bf 69}, 054010 (2004)
  [arXiv:hep-lat/0311028].

 \bibitem{Colangelo:2005gd}
  G.~Colangelo, S.~Durr and C.~Haefeli,
  Nucl.\ Phys.\  B {\bf 721}, 136 (2005)
  [arXiv:hep-lat/0503014].

 \bibitem{Golterman:1997st}
  M.~F.~L.~Golterman and K.~C.~L.~Leung,
  Phys.\ Rev.\  D {\bf 57}, 5703 (1998)
  [arXiv:hep-lat/9711033].

 \bibitem{Brower:2003yx}
  R.~Brower, S.~Chandrasekharan, J.~W.~Negele and U.~J.~Wiese,
  Phys.\ Lett.\  B {\bf 560}, 64 (2003)
  [arXiv:hep-lat/0302005].

 \bibitem{Babich:2006bh}
  R.~Babich, N.~Garron, C.~Hoelbling, J.~Howard, L.~Lellouch and C.~Rebbi,
  Phys.\ Rev.\  D {\bf 74}, 073009 (2006)
  [arXiv:hep-lat/0605016].

\end{thebibliography}
\end{document}